\documentclass[twocolumn,showpacs,preprintnumbers,amsmath,amssymb]{revtex4}

\usepackage{graphicx}
\usepackage{dcolumn}
\usepackage{bm}

\begin{document}

\preprint{APS/123-QED}

\title{Sterile Neutrino-Enhanced Supernova Explosions}

\author{Jun Hidaka}
 \email{jhidaka@ucsd.edu}
\author{George M. Fuller}%
 \email{gfuller@ucsd.edu}
\affiliation{%
Department of Physics, University of California, San Diego, La Jolla, CA 92093-0319
}%



\date{\today}


\begin{abstract}
We investigate the enhancement of lepton number, 
energy, and entropy  transport
resulting from active-sterile neutrino conversion
$\nu_e\rightarrow\nu_s$ deep in the post-bounce supernova core
followed by re-conversion $\nu_s\rightarrow\nu_e$ further out, near the
neutrino sphere. We explicitly take account of shock wave and neutrino
heating modification of the active neutrino forward scattering
potential which governs sterile neutrino production. We find that the $\nu_e$ luminosity 
at the neutrino sphere could be increased by
between $\sim 10\,\%$ and $\sim 100\,\%$ during the crucial shock
re-heating epoch if the sterile neutrino has a rest mass and vacuum
mixing parameters in ranges which include those required for viable
sterile neutrino dark matter. We also find  sterile neutrino transport-enhanced entropy deposition ahead of the shock. This \lq\lq pre-heating\rq\rq\ can help melt heavy nuclei and thereby reduce
the nuclear photo-dissociation burden on the shock. 
Both neutrino luminosity enhancement and pre-heating could increase the likelihood of a successful core collapse supernova explosion.

\end{abstract}


\pacs{14.60.Pq,95.35.+d,97.60.Bw,98.80.-k}
                             Classification Scheme.

\maketitle

\section{Introduction}

There are many purely astrophysical and nuclear physics uncertainties in the core collapse supernova problem. However, the weak interaction in general and neutrino physics in particular play pivotal roles in nearly every aspect of the collapse of the core of a massive star and likely in any subsequent supernova explosion as well. It is sobering to contemplate that collapsing stellar cores will pass through regimes of matter density and neutrino flux which have never been probed in the laboratory and which could be affected significantly by new physics in the weakly interacting sector. Moreover, the existence of neutrino rest masses, unexplained and unpredicted by the Standard Model of particle physics, points directly at the possibility of new neutrino physics.

In this paper we explore the effects of plausible extensions of the Standard Model in the weakly interacting sector on models for the explosion mechanism for core collapse supernovae. In particular, we investigate the effects of an electroweak singlet (\lq\lq sterile\rq\rq) neutrino $\nu_s$ on the physics of energy and lepton number transport in the supernova core and on the process of shock re-heating. The ranges of sterile neutrino rest mass and active-sterile vacuum mixing angle investigated here include those parameters of interest for sterile neutrino dark matter \cite{DM1,DM2,XSF,AFP,DH,AF,2006PhRvD..73f3506A,BiermannKusenko,Kev,AbazajianKoushiappas,1994PhLB..323..360S,x-ray,x-ray_1,Kev2,Boyarsky,Viel,2006PhRvD..74c3009W,2006MNRAS.370..213B} 
and pulsar kicks and related issues \cite{kicks,kicks2,FK}. 
The LSND experiment \cite{LSND,LSND_1,LSND_2}
and recent mini-BooNE experiment \cite{MiniBooNE}
do not constrain the sterile neutrino mass and mixing parameters considered in this paper.

The general features of core collapse supernova evolution are dictated largely by entropy considerations \cite{BBAL}. Stars with initial masses in excess of $\sim 10\,{\rm M}_\odot$ evolve quickly to their evolutionary endpoint: a low entropy core supported by relativistically-degenerate electrons and, therefore, subject to dynamical instability. The collapse of this core is halted at or just beyond the point where nuclear density is reached. The gravitational binding energy released in this prompt collapse and in subsequent quasi-static contraction is more or less efficiently converted into seas of neutrinos of all kinds. The \lq\lq bounce\rq\rq\ of the core generates a shock wave which moves out. However, the energy in this shock is sapped by the photo-dissociation of nuclei passing through it. This process is an inevitable consequence of the substantial entropy jump across the shock front and of basic nuclear physics.

The details of the mechanism or mechanisms whereby the deleterious effects of nuclear photo-dissociation are ameliorated, a viable shock is re-born, and an explosion originates remain elusive. However, ever since the work of Bethe and Wilson \cite{BW85} the broad outlines of a solution are plausibly clear. The prodigious energy in the neutrino and antineutrino reservoirs in the collapsed core is radiated from the surface of the proto-neutron star (the neutrino sphere) and is deposited in material behind the stalled bounce-shock, \lq\lq re-heating\rq\rq\ it and thereby driving a Type II, Ib, or Ic supernova explosion. 

However, one-dimensional simulations of this process, though containing detailed treatments of the nuclear equation of state and neutrino transport, nevertheless are challenged in producing convincing explosions. Much recent attention has focussed on multi-dimensional hydrodynamic, convective, or acoustic enhancement of neutrino energy transport \cite{2006ApJ...640..878B,2006ApJ...642..401B,2002ApJ...574L..65F,2006A&A...453..661K} 
above the neutrino sphere as a means of augmenting neutrino heating of matter below the shock. These schemes succeed in producing explosions. However, as yet they do not include the level of sophistication in, {\it e.g.,} neutrino transport and nuclear equation of state employed in the one-dimensional models for all relevant regimes of time and space. 

Our previous work \cite{Hidaka-Fuller} on the effects of active-sterile-active ($\nu_e \rightarrow \nu_s \rightarrow \nu_e$) neutrino flavor transformation in the in-fall epoch of supernova core collapse suggested a means by which neutrino energy transport could be augmented. Conceivably, this could be a solution to the shock re-heating problem. However, a key uncertainty not addressed in Ref.~\cite{Hidaka-Fuller} was the effect on this process of the shock wave itself. Here we will tackle this issue. 

In Section II we summarize the salient features of active-sterile-active neutrino flavor transformation physics and its effects during the in-fall epoch. In Section III we consider the ways in which the shock wave modifies the thermodynamic conditions which help determine how sterile neutrino production and re-conversion proceed. We also discuss sterile neutrino induced \lq\lq pre-heating\rq\rq\ and the possibility of a reduced nuclear photo-dissociation burden on the shock. In Section IV we discuss shock re-heating and the enhanced prospects for a supernova explosion which could be a by-product of active-sterile-active neutrino conversion schemes. We give conclusions in Section V.

\section{In-Fall Phase Neutrino Flavor Conversion}
In this section we briefly summarize  our previous work \cite{Hidaka-Fuller} 
on the effects of active-sterile neutrino flavor conversion on the in-fall
phase of a core collapse supernova. The key result of
this earlier work was the discovery that electron neutrino conversion 
into a sterile neutrino species $\nu_e \rightarrow \nu_s$ 
could feed back on electron capture ($e^-+p\rightarrow n+\nu_e$) during collapse 
and alter the potential governing
flavor transformation so as to produce a double 
Mikeyev-Smirnov-Wolfenstein (MSW) resonance \cite{MSW,MSW_1}. It is this double resonance structure which can lead to the the re-conversion of the sterile neutrinos. With such a double resonance arrangement, at least some electron neutrinos will experience $\nu_e \rightarrow \nu_s \rightarrow \nu_e$ as they move from higher toward lower density in the core.

For simplicity, we consider $2\times 2$ neutrino flavor mixing where, in vacuum, we have
\begin{eqnarray}
\vert\nu_e\rangle&  = & \cos\theta \vert\nu_1\rangle + \sin\theta\vert\nu_2\rangle ,\\
\label{mix1}
\vert\nu_s\rangle & = & -\sin\theta \vert\nu_1\rangle + \cos\theta\vert\nu_2\rangle .
\label{mix12}
\end{eqnarray}
Here $\theta$ is an effective $2\times 2$ vacuum mixing angle for the $\nu_e\rightleftharpoons\nu_s$ channel, and $\vert\nu_1\rangle$ and $\vert\nu_2\rangle$ are light and heavy, respectively, neutrino energy (mass) eigenstates with mass eigenvalues $m_1$ and $m_2$, respectively. The relevant mass-squared difference is $\delta m^2 \equiv m_2^2-m_1^2$. Since we will be concerned with sterile neutrino rest mass scales $\sim {\rm keV}$, we will have $m_2 \gg m_1$, and so $\delta m^2 \approx m_2^2 \equiv m_{\rm s}^2$.

An electron neutrino ($\nu_e$) propagating coherently in the medium of the core will experience a potential stemming from forward scattering on all particles (electrons/positrons, nucleons/quarks, and other neutrinos) that carry weak charge. This potential is 
\begin{equation}
V  = {{3\sqrt{2}}\over{2}} G_{\rm F} n_{\rm b} \left(Y_e - {{1}\over{3}}+{{4}\over{3}} Y_{\nu_e} + {{2}\over{3}} Y_{\nu_\mu} +  {{2}\over{3}} Y_{\nu_\tau} \right),
\label{V}
\end{equation}
where $n_{\rm b} = \rho N_A$ is the baryon number density, $\rho$ is the density in ${\rm g}\,{\rm cm}^{-3}$  and $N_A$ is Avogadro's number, $G_{\rm F}$ is the Fermi constant, and the net lepton abundances relative to baryons are, {\it e.g.,} $Y_e \equiv {\left( n_{e^-}-n_{e^+}\right)}/n_{\rm b}$ with, {\it e.g.,} $n_{e^-}$ the electron number density. The terms proportional to $Y_{\nu_e}$, $Y_{\nu_\mu}$, and $Y_{\nu_\tau}$ in this potential stem from neutrino-neutrino forward scattering and must be corrected for the non-isotropic nature of the neutrino distribution functions at locations 
which are above the neutrino sphere \cite{Fuller87}. At any point inside the star or above it, electron antineutrinos, {\it i.e.,} $\bar\nu_e$'s, will experience a potential with the same magnitude as that experienced by $\nu_e$'s, but with opposite sign. Of course, sterile neutrinos $\nu_s$ experience no forward scattering potential.
\begin{figure}[htbp]
\includegraphics[width=3.4in]{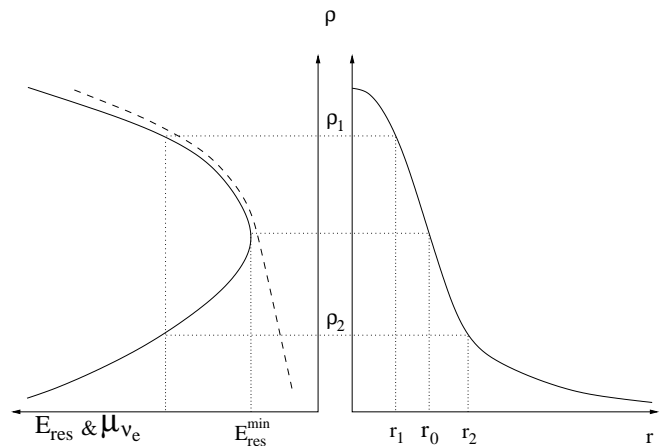}
\caption{\label{fig:E_resAndDensityProfile}Right panel shows the
core density profile with radius $r$, while the corresponding profiles for MSW resonance energy $E_{\rm res}$ (solid) and $\nu_e$ chemical potential
$\mu_{\nu_e}$ (dashed) are shown in the left panel. Here
$E_{\rm res}$ takes its minimum value $E_{\rm res}^{\rm min}$ at
$r_0$. For a particular neutrino energy, an MSW resonance can occur at two locations (densities), {\it e.g.}, $r_1$
($\rho_1$) and $r_2$ ($\rho_2$). }
\end{figure}

At a given location, a neutrino ($\nu_e$ or $\nu_s$) with energy $E_{\rm res}$ will experience an MSW medium-enhanced resonance where
\begin{equation}
E_{\rm res} = {{\delta m^2 \cos2\theta}\over{ 2 V}}\approx  {{m_{\rm s}^2}\over{ 2 V}}.
\label{rescond}
\end{equation}
Physically, this is the neutrino energy where the effective in-medium mass associated with the active neutrino matches the rest mass associated with the sterile state, $m_{\rm s}$. The last approximation in Eq.~(4) follows for the reasons given above and because the vacuum mixing angles we consider here are very small ({\it e.g.,} satisfying $\sin^22\theta \sim {10}^{-9}$).

In medium the forward scattering potential will modify not only the effective masses of the active neutrinos but also the unitary relation between the neutrino flavor states (weak interaction eigenstates) and the (instantaneous) mass eigenstates $\vert \nu_1(t)\rangle$ and $\vert\nu_2(t)\rangle$, where $t$ represents any Affine parameter along the neutrino's world line. We can express the in-medium transformation in direct analogy to that in vacuum,
\begin{eqnarray}
\vert\nu_e\rangle&  = & \cos\theta_{\rm M}(t)\vert\nu_1(t)\rangle + \sin\theta_{\rm M}(t)\vert\nu_2(t)\rangle ,\\
\label{medmix1}
\vert\nu_s\rangle & = & -\sin\theta_{\rm M}(t) \vert\nu_1(t)\rangle + \cos\theta_{\rm M}(t)\vert\nu_2(t)\rangle .
\label{medmix2}
\end{eqnarray}
A similar unitary transformation applies to the antineutrinos but with a different mixing angle $\bar\theta_{\rm M}(t)$.
In an active-sterile neutrino oscillation scenario where neutrino transformation is enhanced and antineutrino transformation is suppressed, at resonance we will have $\theta_{\rm M}(t_{\rm res}) = \pi/4$, {\it i.e.,} maximal mixing. The region in space where the effective in-medium mixing angles $\theta_{\rm M}$ (or $\bar\theta_{\rm M}$) are large and near maximal is termed the resonance width. This width is $\delta t \approx {\vert d \ln V/dt \vert}^{-1}\,\tan2\theta$ and so is expected to be small for the neutrino parameters and conditions we treat here.

So long as the the neutrino mean free paths are large compared to the MSW resonance width, we can regard neutrino flavor evolution as coherent, at least as far as the application of the MSW formalism is concerned \cite{Fuller87}. This is true even when the active neutrinos are trapped and thermalized in the core. Note, however, that at very high densities, such as those we expect to encounter deep in the core near and after core bounce, this condition will break down. There may be so many scattering targets for the active neutrinos in this case that the neutrino mean free paths are comparable to or shorter than the MSW resonance widths. We term this the incoherent or scattering-dominated case. In this regime, scattering-induced de-coherence of the neutrino fields will dominate the conversion of neutrino flavors. In particular, this can be the case for the $\nu_e \rightarrow \nu_s$ channel of most interest here. Note, however, that since the de-coherent neutrino (antineutrino) flavor conversion rate is proportional to $\sin^22\theta_{\rm M}(t)$ ($\sin^22\bar\theta_{\rm M}(t)$), the potential $V$ and the MSW resonance condition still play a significant role in determining the locations where this conversion is significant. Ref.~\cite{AFP} and references therein discuss this physics in detail, while Ref.~\cite{2007PhRvD..75h5004B,2007arXiv0705.0703B} 
discusses uncertainties and controversies associated with de-coherence in high density matter.

Employing a simple nuclear liquid drop model \cite{BBAL,Fuller82} and degenerate electron equation of state in a one-zone homologous collapse code \cite{Fuller82}, we found the double resonance structure
discussed above. 
Fig.~\ref{fig:E_resAndDensityProfile} gives a graphic summary of these results. The equation of state and one-zone collapse code employed in obtaining these results is discussed in the Appendix of Ref.~\cite{Hidaka-Fuller}.

These calculations also showed that near the surface of the core, where the
density is $\rho\sim 10^{12}\,{\rm g/cm^3}$, the MSW resonance energy
$E_{\rm res}$  for $\nu_e\rightleftharpoons\nu_s$ tends to be much larger than
the $\nu_e$ chemical potential (Fermi energy) $\mu_{\nu_e}$.  Progressing inward from the edge of the collapsing core, $E_{\rm res}$ first
decreases while $\mu_{\nu_e}$ increases continuously. Near the density $\rho\sim
10^{13}\,{\rm g/cm^3}$, $E_{\rm res}$ and $\mu_{\nu_e}$ become comparable and
large-scale $\nu_e\rightarrow\nu_s$ conversion starts. Once this conversion process 
begins in earnest,
$E_{\rm res}$ increases with further increases in density. 
In this latter phase of the collapse, $E_{\rm res}$ stays slightly above
$\mu_{\nu_e}$, with both quantities increasing with increasing density. As will be discussed in the next section, a feedback process keeps
$E_{\rm res}$ hovering just above $\mu_{\nu_e}$.
Ultimately,  at core bounce, when the collapse is halted,
the matter density is near nuclear matter density ($\rho\sim 10^{14}\,{\rm g/cm^3}$)
and the relevant neutrino energies are large since $\mu_{\nu_e}\sim 150\,{\rm MeV}$.
Fig.~\ref{fig:E_resAndDensityProfile} illustrates these trends.

Our earlier work \cite{Hidaka-Fuller} speculated that the double MSW resonance structure could facilitate enhanced neutrino energy, entropy, and lepton number transport from deep in the core to regions nearer the proto-neutron star surface ({\it i.e.,} the neutrino sphere). Essentially this enhancement comes about because a neutrino, initially a $\nu_e$ in our case, will spend part of its time as a sterile neutrino. While it is in the sterile state, this neutrino will move at almost the speed of light. As a result, the effective mean free paths and diffusion coefficients for these neutrinos will be re-normalized upward. Interestingly, our estimates suggested that the best prospects for transport enhancement through this mechanism could be obtained with sterile neutrino mass and vacuum flavor mixing parameters which overlap the ranges of these that give viable sterile neutrino dark matter \cite{DM1,DM2,XSF,AFP,DH,AF,2006PhRvD..73f3506A,BiermannKusenko,Kev,AbazajianKoushiappas,1994PhLB..323..360S,x-ray,x-ray_1,Kev2,Boyarsky,Viel,2006PhRvD..74c3009W,2006MNRAS.370..213B}.

However, a significant caveat on these conclusions is that the calculations of Ref.~\cite{Hidaka-Fuller} dealt only with the in-fall epoch of core evolution. The bounce shock generated near the edge of the homologous core could be expected to move outward, through the outer core, and modify the thermodynamic variables and composition in this region. These modifications, in turn, could be expected to alter the $\nu_e$ forward scattering potential which governs sterile neutrino production and/or re-conversion.

\section{Effect of Shock Wave Passage}\label{sec:shockwave_inner}

Assessing the impact of post-shock active-sterile-active neutrino flavor transformation requires adroit attention to a few key issues in supernova shock formation and propagation. As the initial iron core collapses, an inner, homologous core will maintain a roughly self-similar, index 3 polytropic structure \cite{Goldreich-Weber,BBAL}. This makes intuitive sense because the pressure support in the star is dominated by relativistically degenerate electrons with Fermi level (chemical potential) $\mu_e \approx 11.1\,{\rm MeV} {\left( \rho_{10}\, Y_e\right)}^{1/3}$, where $\rho_{10}$ is the density in units of ${10}^{10}\,{\rm g}\,{\rm cm}^{-3}$.  

However, as electron capture proceeds and the pressure is relatively reduced, only a smaller, \lq\lq inner core\rq\rq\ can continue to collapse in this self similar and homologous manner. Homology (in-fall velocity proportional to radius) allows a one-zone calculation to be meaningful, as each location in the inner core will experience a portion of a common temperature, density, and composition history \cite{BBAL}. 

The remainder of the initial iron core which is above and outside the inner core is termed the \lq\lq outer core.\rq\rq\ The inner core is essentially an instantaneous Chandrasekhar mass $M_{\rm IC} \sim \langle Y_e\rangle^2$. When the central density reaches the point where nucleons touch (nuclear density), this core will bounce as a unit and serve as a piston. The shock will form at the edge of this inner core. The initial shock energy will be of order the gravitational binding energy of the inner core and will scale as $~\langle Y_e\rangle^{10/3}$ \cite{Fuller82}. As a result, there is some uncertainty in this initial shock strength depending on nuclear and sub-nuclear density equation of state, composition, and electron capture physics issues. In broad brush, however, we expect the entropy-per-baryon $S$ (in units of Boltzmann's constant $k_{\rm B}$) to jump by a few units at the shock front. 

This entropy jump can be significant because the core's material during the collapse itself, as well as the un-shocked material in the outer core ahead of the shock, is characterized by low entropy, $S \approx 1$. In the lower density regions of the outer core, an entropy jump $\Delta S \ge 3$, for example, is usually enough to shift the nuclear composition in Nuclear Statistical Equilibrium (NSE) from heavy nuclei to free nucleons and alpha particles. We will refer to this phenomenon as nuclear photo-dissociation or nuclear \lq\lq melting.\rq\rq\ 

As the shock propagates through the outer core and melts nuclei it loses energy. This is because each nucleon is bound in a nucleus by $\sim 8\,{\rm MeV}$. This represents ${10}^{51}\,{\rm ergs}$ ($\equiv 1\,{\rm Bethe}$) per $0.1\,{\rm M}_\odot$ of material transiting the shock front. Since the shock is born with an energy $\sim 1\,{\rm Bethe}$ and the outer core mass may be $\sim 0.7\,{\rm M}_\odot$, nuclear photo-dissociation quickly degrades the shock into a \lq\lq dead,\rq\rq\ standing accretion shock.

Whether subsequently the shock can be re-energized by, {\it e.g.,} direct or convectively- or hydrodynamically-enhanced neutrino heating or electromagnetic or acoustic energy transport remains an open question as discussed in the Introduction \cite{2006ApJ...640..878B,2006ApJ...642..401B,2002ApJ...574L..65F,2006A&A...453..661K}. 
However, by any objective standard, the energy ($\sim 1\,{\rm Bethe}$) in observed Type II supernova shocks/explosions is small compared to the energy ($\sim 10\,{\rm Bethe}$) in the neutrino seas initially trapped in the core, and miniscule compared to the energy ($\sim 100\,{\rm Bethe}$) in the neutrino seas a few seconds post-core-bounce. Active-sterile neutrino transformation can tap into this reservoir and change the way in which neutrino energy is transported in and around the supernova core.

As discussed in the last section, direct active-sterile-active neutrino flavor transformation could re-normalize upward the neutrino energy transport rate, thereby increasing the neutrino luminosity at the neutrino sphere and so boosting the shock re-heating rate. Also, the efficacy of the various re-heating schemes may depend on how far out the shock progresses before it stalls. In turn, this depends, among other variables, on electron capture and the shock energy remaining after nuclear photo-dissociation in the outer core. (See the discussion on this point in Ref.~\cite{Hix}.) Any effect like pre-heating which diminishes the nuclear photo-dissociation burden could translate into a larger stall radius for the shock, in turn, helping to increase the effectiveness of the various shock re-heating processes.  

\subsection{Feedback between resonance energy and $\nu_e$ Fermi level}

An important finding in the calculations of Ref.~\cite{Hidaka-Fuller} was that
the active-sterile MSW resonance energy $E_{\rm res}$ exhibited a minimum which was located well inside the core. The density profile and $E_{\rm res}$ profile at bounce is illustrated in Fig.~\ref{fig:E_resAndDensityProfile}. The location of the minimum in $E_{\rm res}$ at bounce is another way to divide the core. As a consequence of this minimum, the first resonance $\nu_e\rightarrow\nu_s$ may occur in the inner core, while the re-conversion resonance, the second one, $\nu_s\rightarrow\nu_e$, typically occurs in the outer core. Note that at the inner resonance, inside of the location of the minimum in $E_{\rm res}$, the $\nu_e$ Fermi energy $\mu_{\nu_e} \approx 11.1\,{\rm MeV} {\left( 2 \rho_{10}\, Y_{\nu_e}\right)}^{1/3}$ tracks just below $E_{\rm res}$, increasing with increasing density just as does $E_{\rm res}$. 

Another key finding of Ref.~\cite{Hidaka-Fuller} was that in the region inside of the resonance energy minimum there is a feedback between sterile neutrino production, $Y_{\nu_e}$, and $Y_e$ which keeps $E_{\rm res}$ tracking just above $\mu_{\nu_e}$. This feedback process is a result of the high degeneracy in the electron neutrino distribution function. If the system were perturbed so that $E_{\rm res}$ were lower than $\mu_{\nu_e}$, there would be prodigious sterile neutrino production which would tend to lower the local net electron lepton number and return the system to a state with $E_{\rm res} > \mu_{\nu_e}$.

\subsection{Shock wave modification of sterile neutrino production}
The passage of the shock through a region can alter the relation between $E_{\rm res}$ and $\mu_{\nu_e}$ and so can influence sterile neutrino production there. As long as
$E_{\rm res}$ stays well above the electron neutrino Fermi energy $\mu_{\nu_e}$, the
production of sterile neutrinos is negligible. However, the shock
wave can supply heat/entropy and can cause a discontinuous
change of physical quantities ({\it e.g.}, density and entropy).
Immediately behind the shock front, we might expect the density
jump to result in a smaller 
gap between $E_{\rm res}$ and $\mu_{\nu_e}$. This could be accompanied by 
enhanced $\nu_s$ production. However, as outlined above, 
we expect this condition to be temporary, as
the feedback effect will push  $E_{\rm
res}$ above $\mu_{\nu_e}$ again.

To take into acount this effect in our one-zone calculation, 
we added heat and entropy \lq\lq by hand.\rq\rq\ Specifically, to simulate
the conditions in newly shocked regions of the core, we instantaneously increased
the density by $\Delta \rho = 10^{13} \,{\rm g/cm^3}$ and 
the entropy-per-baryon (in units of $k_{\rm B}$) 
in three different cases by $\Delta S\sim 0.6,\ 2,\ 3$ 
as measured at density $\rho={10}^{13}\,{\rm g/cm^3}$.
We assume that $\beta$-equilibrium and Nuclear Statistical
Equilibrium (NSE) are attained instantaneously. 
This will be a decent approximation in the very high 
density regions where the first MSW resonance will be 
located, {\it e.g.,} inside or just outside the inner core. 

The entropy increments $\Delta S$ that we employ are chosen 
to be values characteristic of the
early stages of shockwave formation. 
These values are smaller than the 
$\Delta S\sim 10$ entropy jump across the shock
which is expected at later times or larger radius. However, our values make sense in a rough, physical sense: For a Chandrasekhar mass initial iron core ($\sim {10}^{57}$ baryons) collapsing to nuclear saturation density, we expect an in-fall kinetic energy at bounce $\sim {10}^{51}\,{\rm erg}$ which, if dissipated as heat at temperature $T\sim 1\,{\rm MeV}$, would give $\Delta S\sim 1$. (See the discussions in Ref.~\cite{BBAL} and Ref.~\cite{Fuller82}.) Going beyond this crude estimate is tricky.

As best we can ascertain, 
our values of $\Delta S$ at relevant locations 
and epochs in the core bracket the results of some published 
large-scale and detailed numerical simulations.
Both Ref.~\cite{2003PhRvL..91t1102H} and Ref.~\cite{2006A&A...447.1049B}
seem to infer values of $\Delta S$ for relevant locations and epochs
which are within the range we consider here. However, as we will see
below, within this range of entropy jump
there can be significant differences in 
$\nu_e\rightarrow\nu_s\rightarrow\nu_e$ effects.

We calculate $\nu_s$ production and the influence 
of this process on the core in the following manner. 
First, we prepare an initial density profile. This is meant to be 
characteristic of the core just prior to core bounce. 
We take this profile to be that of a self-similarly
contracted (homologous) index $n=3$ polytrope with 
central density $\rho_{\rm central} = 3\times
10^{14} \,{\rm g/cm^3}$. We then choose the location
of the shock front on this profile and take the density there as the
initial density when the shock front arrives. We take the 
other initial physical quantities from the results of our
in-fall one-zone calculation at the initial density. We then apply our
increments in density and entropy.
Following a numerical procedure similar to that used to get the initial model,
we use the results of an appropriate one-zone calculation to get the new, post-shock thermodynamic
and lepton number quantities for the given increments
$\Delta \rho$ and $\Delta S$. We use these altered conditions to estimate
the production of sterile neutrinos and the feedback of this process on the potential $V$.
\subsection{Heating of the outer core}
For a given neutrino energy, we can identify the location of the second, outer resonance by using one-zone collapse calculation results for the run of potential $V$ (or, equivalently, $E_{\rm res}$) and the corresponding density profile. In order to assess the effects
of neutrino flavor re-conversion $\nu_s\rightarrow\nu_e$ in this outer region, we need
to estimate  how many $\nu_e$'s are delivered and how much energy is
deposited at the second resonance. 

This can be estimated by assuming
adiabatic neutrino flavor evolution through MSW resonances. (Ref.~\cite{Hidaka-Fuller} discusses why adiabatic evolution is a good approximation here.) In the adiabatic limit we can assume that all $\nu_e$'s contained in neutrino energy range $\Delta E_{\rm res}$, corresponding to the MSW resonance potential width $\delta V$, are converted to sterile neutrinos $\nu_s$. The width of the resonance in radial coordinate is $\delta r = \left( dr/dV\right) \delta V= {\cal{H}} \tan2\theta $. Here the potential (\lq\lq density\rq\rq ) scale height is ${\cal{H}}=\vert d\ln{V}/dr\vert^{-1}$. Another expression for the spatial resonance width is $\delta r  ={\cal{H}} \Delta E_{\rm res}/E_{\rm res}$. Making use of the resonance condition Eq.~(\ref{rescond}), we can express this as
\begin{equation}
\delta r \approx {{2 V^2 \Delta E_{\rm res}}\over{m_s^2}} {\Bigg\vert{{dV}\over{dr}}\Bigg\vert}^{-1}.
\label{rescond2}
\end{equation}
Using this, we can show that the re-conversion rate per baryon for $\nu_s\rightarrow\nu_e$ at the second
resonance is related to the corresponding rate per baryon for $\nu_e\rightarrow\nu_s$ conversion at the
first resonance by 
%
\begin{equation}
{\dot L}_{\nu_s\rightarrow\nu_e}=\frac{r_{\rm 1st}^2\rho_{\rm 1st}}{r_{\rm
2nd}^2\rho_{\rm 2nd}} \frac{dV/dr|_{\rm 2nd}}{dV/dr|_{\rm 1st}}
{\dot L}_{\nu_e\rightarrow\nu_s}.
\label{eq:conversion_rate} 
\end{equation}
At any location we can designate $L \equiv Y_e+Y_{\nu_e}$ as the total electron lepton number per baryon. Neutrino flavor conversion $\nu_e\rightarrow\nu_s$ ($\nu_s\rightarrow\nu_e$) produces a negative (positive) time rate of change of this quantity, ${\dot L}= dL/dt$, respectively. In employing Eq.~(\ref{eq:conversion_rate}), we evaluate $dV/dr$ numerically using the in-fall one-zone calculation profile. In this equation, $\rho_{\rm 1st}$ ($\rho_{\rm 2nd}$)
and $r_{\rm 1st}$ ($r_{\rm 2nd}$) are the density and the location of the first
(second) resonance, respectively, as illustrated schematically in
Fig.~\ref{fig:E_resAndDensityProfile}. The energy transfer rate per baryon from the
first to the second resonance obeys a relationship in obvious analogy to that in Eq.~(\ref{eq:conversion_rate}). 
\begin{figure}[htbp]
\includegraphics[width=3.2in]{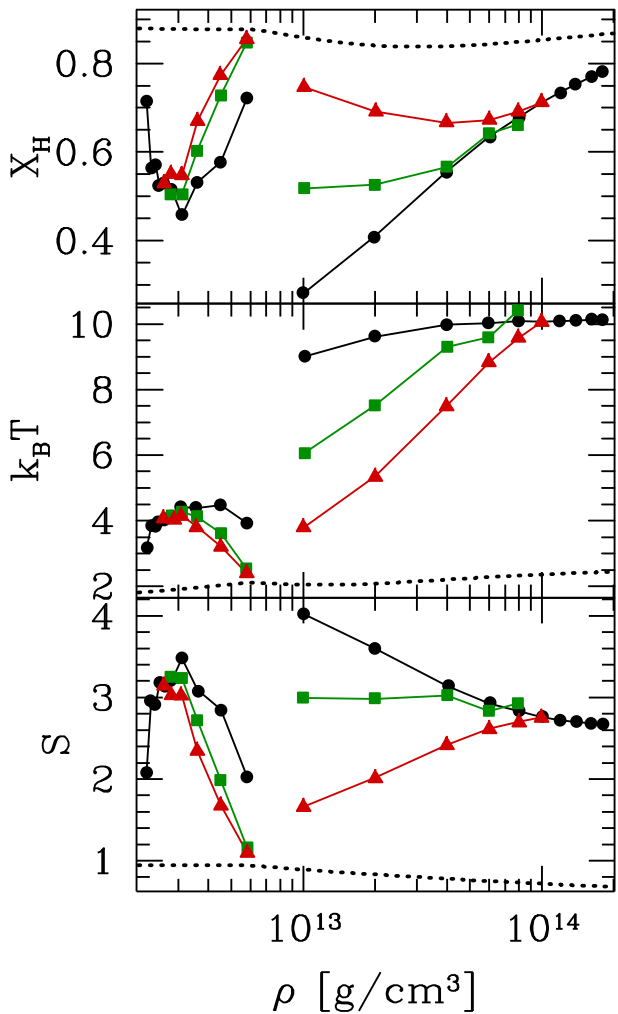}
\caption{\label{fig:ShockEffectHeating}Effects of the shock in the interior (curves on the right) and shock-modified $\nu_e\rightarrow\nu_s\rightarrow\nu_e$ pre-heating in the outer core (curves on the left). Heavy nucleus mass fraction $X_{\rm H}$, temperature $k_{\rm B}
T$ (in MeV), and entropy per baryon $S$ (in units of $k_{\rm B}$) are shown for three different cases. These cases correspond to three different shock strength scenarios with entropy jump
(as measured at $\rho=3\times{10}^{13}\,{\rm g}\,{\rm cm}^{-3}$) $\Delta S =0.6$ (triangles), $\Delta S =2$ (squares), and $\Delta S =3$ (circles), respectively. Results of the pre-shock, in-fall one-zone
calculation are also included (dotted lines). The minimum of the
resonance energy is located around $\rho=7\times 10^{12}\,
{\rm g}\,{\rm cm}^{-3}$. This location divides the curves on the left and right, as described in the text. }
\end{figure}

At the location of the second, outer resonance we take account of the heat and lepton number deposited by $\nu_s\rightarrow \nu_e$ by re-running the one-zone code with these updated quantities but with the density fixed at its original value. This gives us estimates of the change in thermodynamic variables that accompany this \lq\lq pre-heating.\rq\rq\
We continue this calculation of $\nu_e\rightarrow\nu_s\rightarrow\nu_e$
energy transfer to locations in the outer core until the shock
wave reaches the position $r_{0}$ where $E_{\rm res}$ takes its
minimum value (see
Fig.~\ref{fig:E_resAndDensityProfile}). 

Fig.~\ref{fig:ShockEffectHeating}
shows the profiles of entropy, temperature, and heavy nucleus mass
fraction $X_{\rm H}$ at the completion of this energy transfer process. Three profiles
are shown, corresponding to three different shock strength scenarios with entropy jump
(as measured at $\rho=3\times{10}^{13}\,{\rm g}\,{\rm cm}^{-3}$) $\Delta S =0.6$ (triangles), $\Delta S =2$ (squares), and $\Delta S =3$ (circles), respectively.
We may view
these profiles as snapshots of conditions when the shock front is located at
$r_{0}$.  The figure also includes the results of the original
in-fall calculation for comparison.
%
Fig.~\ref{fig:ShockEffectHeating} shows that, depending on shock strength and the $E_{\rm res}$ profile, sterile neutrino-induced pre-heating could result in at least partial ($\sim 50\%$) melting of heavy nuclei in the outer regions of the core ahead of the shock. This could represent a substantial reduction in the nuclear photo-dissociation burden for the shock. Even though our estimates are schematic in nature and crude on a quantitative level, this result is sufficiently dramatic that it is clear that the existence of sterile neutrinos in the mass and mixing ranges discussed here could alter the the energetics of core collapse supernova shock propagation. 
 
%
\section{Shock Re-Heating}
In their pioneering work on core collapse supernovae, Mayle and Wilson \cite{1988ApJ...334..909M,1993PhR...227...97W} obtained vigorous explosions in the late-time shock re-heating model, even in one dimension. This result was, and continues to be, at odds with the results of other detailed one-dimensional simulations, some more sophisticated in their treatments of the nuclear equation of state and neutrino transport \cite{2004ApJS..150..263L,2003ApJ...592..434T,2005ApJ...620..840L,2005AAS...207.1701M,2006ApJ...642..401B,2004rpao.conf..224L,2005ApJ...626..317W,2002ApJ...574L..65F,2005PhRvD..72d3007C,2001ApJ...560..326B,2006astro.ph..7281S}. 
Mayle and Wilson got their result by invoking neutrino convective transport in the core to increase the neutrino luminosity at the neutrino sphere. Though the physical basis for this effect ({\it i.e.,} their \lq\lq neutron fingers\rq\rq) has been repudiated, their result taught us a valuable lesson: The efficacy of neutrino heating in re-enegizing the stalled shock is a sensitive function of neutrino and antineutrino transport in the core and the corresponding luminosities at the neutrino sphere. The process of $\nu_e\rightarrow\nu_s\rightarrow\nu_e$ flavor conversion in the core could be just the sort of neutrino energy transport augmentation that could aid the core collapse supernova explosion process \cite{Hidaka-Fuller}.
%

%
\subsection{De-coherent production of sterile neutrinos inside the proto-neutron star}
Neutrino flavor evolution deep in the central region of the post-bounce core will be collisionally-dominted. The characteristic density in the central core at this epoch will be near or above nuclear saturation density, $\rho\sim 3\times{10}^{14}\,{\rm g}\,{\rm cm}^{-3}$, and scattering-induced de-coherence will be the primary channel through which sterile neutrinos are produced from the seas of active neutrinos \cite{2001PhRvD..64b3501A}.

The total (left-handed $\nu_s$ plus right-handed $\bar\nu_s$) sterile neutrino emissivity $\mathcal{E}$ (energy emission per unit mass per unit time) can be estimated by employing average neutrino and  antineutrino flavor conversion probabilities $\langle
P_m(\nu_{e}\rightarrow\nu_{s};p,t)\rangle$ and $\langle
P_m(\bar{\nu}_{e}\rightarrow\bar{\nu}_{s};p,t)\rangle$, respectively, as functions of neutrino or antineutrino momentum $p$ and location parameter $t$, energy-dependent neutrino and antineutrino scattering cross sections (in principle on all weakly interacting targets) $\sigma_{\nu_{s}}(E)$ and $\sigma_{\bar{\nu}_{e}}(E)$, respectively, and integrating over neutrino and antineutrino fluxes and energies $E$ \cite{1991NuPhB.358..435K,1993APh.....1..165R,1996slfp.book.....R,2001PhRvD..64b3501A},

%
\begin{eqnarray}
\mathcal{E}&\approx& \frac{1}{m_N}\int d\Phi_{\nu_{e}} E
\sigma_{\nu_{s}}(E)\frac{1}{2}\langle
P_m(\nu_{e}\rightarrow\nu_{s};p,t)\rangle \nonumber \\ &+&
\frac{1}{m_N}\int d\Phi_{\bar{\nu}_{e}} E
\sigma_{\bar{\nu}_{e}}(E)\frac{1}{2}\langle
P_m(\bar{\nu}_{e}\rightarrow\bar{\nu}_{s};p,t)\rangle,
\label{eq:emissivity}
\end{eqnarray}
where $m_N$ is an atomic mass unit (essentially, the average free nucleon mass).
In the conditions of near weak and near thermal equilibrium in the post-bounce central core, the differential neutrino and antineutrino fluxes $d\Phi_{\nu_{e}}$
and $d\Phi_{\bar{\nu}_{e}}$ (or number densities $dn_{\nu_e}$ and $dn_{\bar\nu_e}$), respectively, can be expressed as
\begin{eqnarray}
d\Phi_{\nu_{e}}=c dn_{\nu_{e}} &\approx&
\frac{d^3p}{(2\pi)^3}\frac{1}{e^{E/T_{\nu_{e}}-\eta_{\nu_{e}}}+1}\nonumber\\
&\approx&
\frac{1}{(2\pi)^3}\frac{E^2dE}{e^{E/T_{\nu_{e}}-\eta_{\nu_{e}}}+1},
\end{eqnarray}
\begin{eqnarray}
d\Phi_{\bar{\nu}_{e}}=c dn_{\bar{\nu}_{e}} &\approx&
\frac{d^3p}{(2\pi)^3}\frac{1}{e^{E/T_{\bar{\nu}_{e}}-\eta_{\bar{\nu}_{e}}}+1}\nonumber\\
&\approx&
\frac{1}{(2\pi)^3}\frac{E^2dE}{e^{E/T_{\bar{\nu}_{e}}-\eta_{\bar{\nu}_{e}}}+1},
\end{eqnarray}
where the $\nu_e$ ($\bar\nu_e$) degeneracy parameter is
$\eta_{\nu_{e}}=\mu_{\nu_{e}}/T_{\nu_{e}}$ ($\eta_{\bar\nu_{e}}=\mu_{\bar\nu_{e}}/T_{\bar\nu_{e}}\approx -\eta_{\nu_{e}}$), respectively.  The neutrino and antineutrino temperatures $T_{\nu_e}$ and $T_{\bar\nu_e}$, respectively, are essentially the same as the matter temperature. Here the speed of light is $c$. The average oscillation (transformation)
probabilities in Eq.~(\ref{eq:emissivity}) are given by
\begin{eqnarray}
\label{P1}
\lefteqn{\langle P_m(\nu_{e}\rightarrow\nu_{s};p,t)\rangle}\nonumber\\
&\approx& \frac{1}{2}\frac{\Delta(E)^2\sin^2
2\theta}{\Delta(E)^2\sin^2 2\theta+D^2+[\Delta(E)\cos
2\theta-V]^2},\nonumber\\ &&
\end{eqnarray}

\begin{eqnarray}
\label{P2}
\lefteqn{\langle
P_m(\bar{\nu}_{e}\rightarrow\bar{\nu}_{s};p,t)\rangle}\nonumber\\
&\approx& \frac{1}{2}\frac{\Delta(E)^2\sin^2
2\theta}{\Delta(E)^2\sin^2 2\theta+\bar{D}^2+[\Delta(E)\cos
2\theta+V]^2}.\nonumber\\ &&
\end{eqnarray}
Following Ref.~\cite{2001PhRvD..64b3501A}, and for the purpose of simple estimation, here we will take the $\nu_e$ and $\bar\nu_e$ scattering cross sections to be those appropriate for free nucleons. These are roughly
\begin{equation}
\sigma_{\nu_{e}}(E)\approx\sigma_{\bar{\nu}_{e}}(E)\approx 1.66 G_{\rm
F}^2 E^2.
\end{equation}
In Eqs.~(\ref{P1}) and (\ref{P2}), we employ the notation
\begin{equation}
\Delta(p)\equiv \delta m^2/2p\approx m_s^2/2E\approx \Delta(E).
\end{equation}
The quantum damping rate for neutrinos is
\begin{equation}
D=\Gamma_{\nu_{e}}/2=\int d\Phi_{\nu_{e}} \sigma_{\nu_{e}}(E)/2.
\end{equation}
%
The analogous quantum damping rate for antineutrinos, $\bar{D}$,
has a form directly analogous to that for $D$.

The effect of the de-coherent $\nu_s$ and $\bar\nu_s$ production on the
potential $V$ has been studied in the context of a collapsed stellar core in Ref.~\cite{2001PhRvD..64b3501A}. There it was argued that $V$ should evolve toward zero
on a time scale short compared to the characteristic proto-neutron star core dynamical time scale. Accordingly, we shall take $V=0$ in Eq.~(\ref{P1}) and Eq.~(\ref{P2}) in the following discussion. This will facilitate a simple estimate of the sterile neutrino emissivity deep in the central region of the proto-neutron star after bounce.
\subsection{Enhancement of neutrino luminosity behind the shock}
We have estimated the effects of shock passage on thermodynamic and composition variables in the outer parts of the core by employing one-zone simulations of shock propagation through these regions. In doing this, we use the same numerical procedure described in Section III for gauging the effects of shock passage in the inner parts of the core. However, in the case of the outer core, we take account of the $\nu_e\rightarrow\nu_s\rightarrow\nu_e$ pre-heating of the material prior to the arrival of the shock. Therefore, our initial conditions for shock passage in the outer core for this calculation are chosen to be those given by the
$\nu_s\rightarrow\nu_e$ energy deposition process described in Section III and shown in Fig.~\ref{fig:ShockEffectHeating}.

\begin{figure}[htbp]
\includegraphics[width=3.2in]{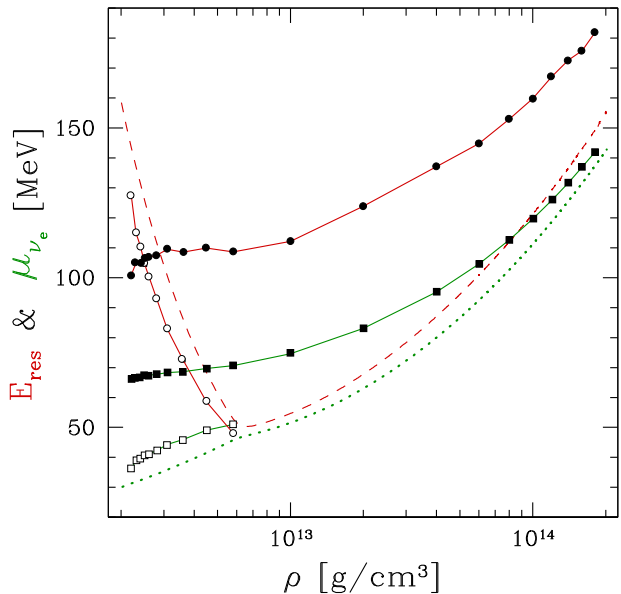}
\caption{\label{fig:ShockEffectInnerOuterEresProfileStrongShockType}
One-zone calculation results for resonance energy $E_{res}$ (in MeV) and $\nu_e$ chemical potential $\mu_{\nu_e}$ (Fermi energy, in MeV) are shown as functions of density $\rho$ (in ${\rm g}\,{\rm cm}^{-3}$). Circles and squares represent $E_{\rm res}$ and $\mu_{\nu_e}$, respectively. Filled
symbols correspond to the values of these quantities for an assumed strong shock ($\Delta S=3$, as described in Section III). This case includes the effect of $\nu_s\rightarrow\nu_e$
reconversion and associated pre-heating ahead of the shock, as well as the effect of the shock itself. The effect of post-bounce pre-heating alone is shown by the quantities with the open circles and squares.
For comparison, $E_{\rm res}$ (dashed line) and $\mu_{\nu_e}$ (dotted line) are given for the in-fall (pre-shock, no pre-heating) case.}
\end{figure}

The results are intriguing. For the case of a strong initial shock ($\Delta S=3$ as measured at density $\rho = {10}^{13}\,{\rm g}\,{\rm cm}^{-3}$), our calculations show that the double resonance structure characteristic of the in-fall regime is destroyed. In this case, however, the resonance energy $E_{\rm res}$ remains well above the $\nu_e$ Fermi energy $\mu_{\nu_e}$. This, in turn, suggests that any $\nu_e$  which is converted to a sterile neutrino $\nu_s$ by scattering-induced de-coherence deep inside the core, yet possesses an energy above the value of $E_{\rm res}$ at the neutrino sphere, will encounter an MSW resonance further out, nearer the neutrino sphere, and will be coherently and adiabatically re-converted to a $\nu_e$ there. Fig.~\ref{fig:ShockEffectInnerOuterEresProfileStrongShockType} shows the results of the one-zone calculations that suggest this scenario. 

\begin{figure}[htbp]
\includegraphics[width=3.2in]{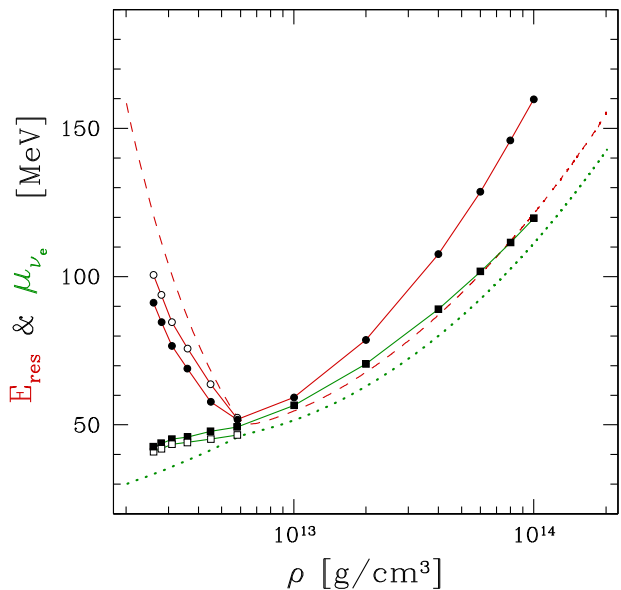}
\caption{\label{fig:ShockEffectInnerOuterEresProfileWeakShockType} Same as Fig.3, but for the case of a weak shock ($\Delta S = 0.6$). }
\end{figure}

It can be seen in Fig.~\ref{fig:ShockEffectInnerOuterEresProfileStrongShockType} that both the $E_{\rm res}$ and $\mu_{\nu_e}$ curves are monotonic with increasing density and each has positive slope. Therefore, the highest energy neutrinos will tend to deposit their energy ({\it i.e.,} be re-converted to $\nu_e$'s) deepest in the core. This could result in more heating by $\nu_e\rightarrow\nu_s\rightarrow\nu_e$ transport enhancement with increasing depth which could, in turn, promote convective instability and further augmentation of neutrino energy transport. 

In any case, since our estimates show that the resonance energy $E_{\rm res}$ asymptotes out to about $E_{\rm res}^{\rm edge} \approx 100\,{\rm MeV}$ at the outer edge of the core, we can conclude that the $\nu_e$'s converted to sterile species in the inner regions of the core where $\mu_{\nu_e} \ge E_{\rm res}^{\rm edge}$ will be reconverted to $\nu_e$'s prior to escaping the core. On account of the quadratic energy dependence of the $\nu_e$ absorption cross sections, such re-converted high energy $\nu_e$'s are certain to deposit their energy and be thermalized on times scales short compared to any transport time scale.

Our calculations suggest that a weaker initial shock will not eliminate the double resonance structure left at the end of the in-fall epoch. Fig.~\ref{fig:ShockEffectInnerOuterEresProfileWeakShockType} is analogous to Fig.~\ref{fig:ShockEffectInnerOuterEresProfileStrongShockType} but shows the results of a one-zone calculation with initial shock strength $\Delta S=0.6$. In this case neither the $\nu_e\rightarrow\nu_s\rightarrow\nu_e$ pre-heating of the outer core or the shock passage event itself can change composition, density, and temperature enough to disrupt the general form of the runs for $E_{\rm res}$ and $\mu_{\nu_e}$. 

We conclude that there may be a threshold in shock strength beyond which the double resonance structure at the end of in-fall is replaced by the single outer resonance regime in Fig.~\ref{fig:ShockEffectInnerOuterEresProfileStrongShockType}. What is this threshold in shock strength? The answer to this question is hard to get at with our simplistic model. However, a fair guess based on our one-zone scheme with its liquid drop equation of state would be $\Delta S \ge 2$ (as measured at $\rho = {10}^{13}\,{\rm g}\,{\rm cm}^{-3}$).

This is significant but, ultimately,  unsatisfying because large-scale numerical supernova simulations, depending on the initial model and on in-fall physics, may produce initial bounce shocks with strengths below, near, or above this threshold. For example, the calculations by the Mezzacappa group \cite{Hix}
, appear to produce shocks with strengths $\Delta S \approx 2$ by our measure. This would be near or above the threshold for erasing the in-fall epoch double resonance structure. However, the simulations by the Janka group \cite{2006A&A...447.1049B}
 suggests a range of shock strengths which could be near the threshold. This issue has to be resolved before we can be confident of the effects of $\nu_e\rightarrow\nu_s\rightarrow\nu_e$ on core collapse supernovae.

Note, however, that in either the weak or strong shock case, sterile neutrinos produced at high energies deep in the core could be converted to $\nu_e$'s further out. This is all we need to enhance energy deposition behind the shock and, therefore, increase the shock re-heating rate. All that remains is an estimate of this heating rate. This requires an estimate of the sterile neutrino emissivity deep in the core.

Following Ref.~\cite{2001PhRvD..64b3501A}, we can get a
rough estimate of the energy radiated in sterile neutrinos $\nu_s$ per unit mass and per unit time \-- the emissivity \-- in the region of the core where the $\nu_e$ Fermi energies are $\mu_{\nu_e} \ge 100\,{\rm MeV}$. As outlined above, we take $V=0$ deep inside the proto-neutron star and approximate the
Fermi distribution as a step function, {\it i.e.,} completely degenerate, with degeneracy parameter $\eta_{\nu_{s}} \gg 1$. In this limit, flavor conversion in the channel $\nu_e\rightarrow\nu_s$ gives rise to sterile neutrino $\nu_s$ emissivity
\begin{eqnarray}
\mathcal{E}&=& \frac{1.66\, G_{\rm F}^2}{8\pi^2 m_N} \sin^22\theta
\int_0^{\mu_{\nu_e}} dE \frac{E^5}{1+{{4D^2 E^2}/{m_{\nu_s}^4}}}
\nonumber\\ &=&\frac{1.66\, G_{\rm F}^2}{16\pi^2 m_N} \sin^22\theta
\left(\frac{m_{\nu_s}^2}{2D}\right)^6\nonumber\\
&&\qquad\times\left[\frac{\xi^4}{2}-\xi^2+\ln (1+\xi^2)\right]
{\Bigg\vert}_{\xi={{2D \mu_{\nu_e}}/{m_{\nu_s}^2}}}
\label{emissivity1}
\end{eqnarray}
where we ignore contributions to the emissivity stemming from $\bar{\nu}_e$'s. Noting that the integration parameter $\xi$ satisfies $\xi\ll 1$
for $100\,{\rm eV}< m_{\nu_s} < 1\,{\rm MeV}$ and that typically
$\mu_{\nu_e}\sim 150\,{\rm MeV}$, we can calculate the emissivity to leading order in $\xi$ to find
%
%
\begin{equation}
\mathcal{E}\approx \left( 2\times 10^{28}\, {\rm erg}\ {\rm s}^{-1}\,{\rm g}^{-1}\right) \sin^22\theta .
\end{equation}
This is then the rate per gram at which energy in sterile neutrinos is flowing out of the inner parts of the core.

On account of adiabatic MSW resonant $\nu_s\rightarrow\nu_e$ flavor conversion, the fraction of the deep core's $\nu_s$ energy flux which is carried by neutrinos with energies above the resonance energy at the outer edge of the core, $E_{\rm res}^{\rm edge}$, will be deposited in the regions just below the neutrino sphere. Using a calculation in obvious analogy to that in Eq.~(\ref{emissivity1}), we can estimate the effective emissivity for this \lq\lq re-captured\rq\rq\ sterile neutrino energy, 


%
\begin{eqnarray}
\lefteqn{\mathcal{E}(\nu_s\rightarrow\nu_e)}\nonumber\\
&\approx&\frac{1}{m_N}\int_{E_{\rm res}^{\rm edge}}^{\mu_{\nu_e}} d\Phi_{\nu_{s}} E
\sigma_{\nu_{e}}(E)\nonumber\\ &&\qquad\times\frac{1}{2}\langle
P_m(\nu_{s}\rightarrow\nu_{e};p,t)\rangle
\end{eqnarray}
where, as argued above,  $E_{\rm res}^{\rm edge} \approx 100\,{\rm MeV}$.
Using the same approximations made in evaluating Eq.~(\ref{emissivity1}), we find
%
\begin{equation}
\mathcal{E}(\nu_s\rightarrow\nu_e)\approx 1.4\times 10^{52}\,{\rm erg}\ {\rm s}^{-1}\,{\rm M}_\odot^{-1}\left(  {{\sin^22\theta}\over{{10}^{-9}}}\right).
\label{re-emit}
\end{equation}
%
%
%
%
Since the inner part of the core which generates the sterile neutrinos has a mass $\sim 1\,{\rm M}_\odot$, the energy deposited per unit time near the edge of the neutron star could be prodigious. Of course, this conclusion depends on a host of active-sterile neutrino mass/mixing matrix issues including, {\it e.g.,} the effective $2\times 2$ angle $\theta$ characterizing $\nu_e\rightleftharpoons\nu_s$ vacuum mixing. 

If we take $m_s \sim 1\,{\rm keV}$ and $\sin^22\theta = {10}^{-9}$, corresponding to the \lq\lq sweet spot\rq\rq\ for sterile neutrino dark matter and beneficial supernova effects picked out in Ref.~\cite{Hidaka-Fuller}, then the emissivity in Eq.~(\ref{re-emit}) suggests that we could possibly {\it double} the $\nu_e$ energy resident just below the neutrino sphere. Though this energy would be deposited in the form of $\nu_e$'s, rapid re-establishment of beta equilibrium would imply that this energy is shared among all six active neutrino species. This energy sharing roughly will be weighted by the relative numbers of active neutrino species in equilibrium. However, if the extra $\nu_e$'s are deposited quite close to neutrino sphere, energy re-distribution becomes a difficult neutrino transport issue. Since there is a preponderance of $\nu_e$'s, we can guess that there will not be equal amounts of energy in the $\nu_e$, $\bar\nu_e$, $\nu_\mu$, $\bar\nu_\mu$, $\nu_\tau$, and $\bar\nu_\tau$ seas.

On the other hand, since shock re-heating is mostly effected through the charged current capture processes $\nu_e+n\rightarrow p+e^-$ and $\bar\nu_e + p\rightarrow n +e^+$, it is the $\nu_e$ and $\bar\nu_e$ luminosities at the neutrino sphere which are most important. We could be conservative and assume equal energy sharing so that the $\nu_e$ and $\bar\nu_e$ seas get a third of the extra energy deposited by $\nu_s\rightarrow\nu_e$ near the neutrino sphere. In this case, and for a range of mixing angles relevant for sterile neutrino dark matter, we could expect roughly a $\sim 10\%$ to $\sim 100\%$ increase in the sum of the $\nu_e$ and $\bar\nu_e$ luminosities. This, in turn, could lead to comparable increases in the re-heating rate of the shock. 
%
%
%
%
%
%

\section{Lepton Number Transport and the Role of $\mu$- and $\tau$- Flavor Neutrinos}

In this section we discuss the active-sterile-active neutrino flavor transformation-induced flows of electron, muon, and tau lepton numbers and the effects of these on supernova physics. The $\nu_e\rightarrow\nu_s\rightarrow\nu_e$ process outlined above will transport electron lepton number from deep in the core to the vicinity of the neutrino sphere. In the course of describing this process, we made no consideration for mu ($\nu_\mu$, $\bar\nu_\mu$) and tau ($\nu_\tau$, $\bar\nu_\tau$) flavor neutrinos. Surely, if electron neutrino flavors mix in vacuum with a sterile species, likely so will mu and tau flavor neutrinos.  

In broad brush, the lepton number transport rate for $\nu_e\rightarrow\nu_s\rightarrow\nu_e$ should dominate over the rate for $\bar\nu_e\rightarrow\bar\nu_s\rightarrow\bar\nu_e$ and, for that matter, the rates for $\nu_\mu\rightarrow\nu_s\rightarrow\nu_\mu$, $\bar\nu_\mu\rightarrow\bar\nu_s\rightarrow\bar\nu_\mu$, $\nu_\tau\rightarrow\nu_s\rightarrow\nu_\tau$, and $\bar\nu_\tau\rightarrow\bar\nu_s\rightarrow\bar\nu_\tau$ as well. The argument to support this assertion is based on the relative populations of the various active neutrino species.

Keep in mind that the inner core, the \lq\lq piston\rq\rq\ for shock generation at bounce, though experiencing an increase in entropy stemming from the dissipation of in-fall kinetic energy, nevertheless remains relatively low in entropy and full of its original electron lepton number excess. Immediately after bounce, the temperature in the core is $T\sim 10\,{\rm MeV}$, while the $\nu_e$ Fermi energy is $\mu_{\nu_e} \sim 100\,{\rm MeV}$. (See the discussion in Ref.~\cite{1999ApJ...513..780P}.)
 In the standard stellar collapse model, these conditions will persist for of order a neutrino diffusion time scale, {\it i.e.,} seconds. This is a time comparable to or longer than the shock re-heating time of interest here.

The $\bar\nu_e$'s will have a negative chemical potential ($-\mu_{\nu_e}$). The mu and tau flavor neutrinos must be pair produced, and as a consequence they will have zero chemical potential. In the conditions of beta equilibrium in the inner core, the number density of $\nu_e$'s will be $\sim {\mu_{\nu_e}^3}$, while the number density of $\bar\nu_e$'s will be $\sim T^3 \exp{\left(-{\mu_{\nu_e}}/T\right)}$, and the number densities of all mu and tau flavor neutrino species will be $\sim T^3$. Clearly, there should be a large excess of $\nu_e$'s over the other neutrino species in the time frame of interest. As a result, during this time, de-coherence associated with the scattering of active neutrino species will produce far more sterile neutrinos ($\nu_s$'s) than the opposite handedness \lq\lq anti\rq\rq-sterile neutrinos ($\bar\nu_s$'s). 



The picture we have of the supernova core in this time frame is then as follows. We have an inner core  \lq\lq source\rq\rq\ producing a large flux of very high energy $\nu_s$'s and lower fluxes of lower energy $\bar\nu_s$'s. The $\nu_s$'s will be preferentially transformed to $\nu_e$'s via $\nu_s\rightarrow\nu_e$ near the neutrino sphere. This is because in this region the forward scattering potential for mu or tau neutrino conversion to sterile neutrinos will be negative. With a negative potential, only antineutrinos can be matter-enhanced.

For example, the forward scattering potential for the flavor conversion channel
$\nu_s\rightleftharpoons\nu_{\mu}$ is given by
\begin{eqnarray}
V_{\nu_\mu} & = & {{\sqrt{2}}\over{2}}\, G_{\rm F}\, n_{\rm b} (
Y_e-1+2Y_{\nu_e}\nonumber\\
&&\qquad\qquad+4Y_{\nu_{\mu}}+2Y_{\nu_{\tau}} ).
\label{mupot}
\end{eqnarray}
An analogous expression holds for the potential, $V_{\nu_\tau}$, relevant for $\nu_s\rightleftharpoons\nu_\tau$, but with the coefficients of $Y_{\nu_\mu}$ and $Y_{\nu_\tau}$ in Eq.~(\ref{mupot}) swapped.
Since we expect $Y_e \approx 0.35$, $Y_{\nu_e} \sim 0.05$, and $Y_{\nu_\mu} = Y_{\nu_\tau}=0$ initially, we will have $V_{\nu_\mu} < 0$ and $V_{\nu_\tau} < 0$. This, in turn, implies that only the channels $\bar\nu_s\rightleftharpoons\bar\nu_\mu$ and $\bar\nu_s\rightleftharpoons\bar\nu_\tau$, respectively, can be matter-enhanced and resonant near the neutrino sphere. These processes will be sub-dominant compared to $\nu_s\rightleftharpoons\nu_e$ because the energy and fluxes for the $\bar\nu_s$'s will be lower than those for $\nu_s$'s as argued above.

The dominant $\nu_s\rightarrow\nu_e$ conversion process will lead to the region near the neutrino sphere being \lq\lq charged up\rq\rq\ with positive electron lepton number. Given the energy emissivities discussed in the last section, for example we might expect an additional electron lepton number per baryon $\Delta Y_{\nu_e} \sim {10}^{52}\,{\rm erg}\,{\rm s}^{-1}/\left( 100\,{\rm MeV}\cdot {10}^{57}\ {\rm baryons}\right) \approx 0.1$ to be deposited over a time $\sim 1\,{\rm s}$ after core. (Likewise, there will be a corresponding, though far smaller increase in negative mu and/or tau lepton number stemming from $\bar\nu_s\rightarrow\bar\nu_{\mu,\tau}$.)

The $\nu_e$'s deposited by $\nu_s\rightarrow\nu_e$ could represent a significant increase in electron lepton number. In turn, this will tend to decrease the neutron excess. This is because the additional $\nu_e$'s will tend to shift the equilibrium relation, $\nu_e+n\rightleftharpoons p+e^-$, to the right, producing higher electron fraction $Y_e$ and more protons. Likewise, the neutron-to-proton ratio, $n/p= Y_e^{-1}-1$, in the material near the neutrino sphere will be transmitted by the $\nu_e$ and $\bar\nu_e$ fluxes emergent from the neutrino sphere to the material in the region between the neutron star and the shock \cite{Qian}.

In the early shock re-heating regime, this increase in electron
fraction $Y_e$ in the material ejected by neutrino heating could be
beneficial for nucleosynthesis. In the calculations of nucleosynthesis
in early, shock re-heating epoch neutrino-heated ejecta performed by
Woosley {\it et al.} using the Mayle and Wilson supernova simulation
results \cite{1994ApJ...433..229W}, it was found that there was an
overproduction of neutron number $N=50$ nuclei. Subsequently it was
pointed out in Ref.~\cite{1996ApJ...460..478H} that a modest increase
in $Y_e$ could cure this problem. The
$\nu_e\rightarrow\nu_s\rightarrow\nu_e$ lepton number transfer process
at least sends $Y_e$ in the right direction at the right epoch to
help. 

Effects of active-sterile-active neutrino flavor transformation
at later times, in the post-shock revival hot bubble, may be very
interesting, but are beyond the scope of the current work. 

We note that re-conversion of sterile neutrinos has been considered
previously in models for r-process nucleosynthesis
\cite{1999PhRvC..59.2873M,Fetter:2002xx}. These calculations, however,
concentrated on the late-time regime above the core and considered a
much different sterile neutrino mass and vacuum mixing range from the
one considered here. Additionally, active-active neutrino flavor
transformation in the supernova environment is a very difficult
problem
\cite{Fuller87,1988NuPhB.307..924N,1992PhRvD..46..510P,1993NuPhB.406..423S,FM,Qian,PastorRaffelt,2005NJPh....7...51B,2006PhRvD..73b3004F,2006PhRvD..74j5010H,2006PhRvD..74l3004D,2006PhRvD..74j5014D,2006PhRvL..97x1101D,duan:125005}. A
complete assessment of nucleosynthesis effects would neccssitate
treating all active-active and active-sterile neutrino flavor
conversion processes.


\section{Conclusions}

Generally it has been assumed that the emission of sterile neutrinos from the supernova core will tend to decrease the prospects for obtaining a successful core collapse supernova explosion. This may be true if a large enough amount of energy is lost from the core. This is because, after all, most of the gravitational binding energy released in the collapse of the core and subsequent quasi-static contraction of the hot proto-neutron star is \lq\lq stored\rq\rq\ in trapped seas of active neutrinos of all species. Moreover, it is this neutrino energy which, ultimately, will be invoked one way or another to revive the nuclear photo-dissociation-degraded bounce shock. 

However, in this paper we point out that the notion that sterile neutrino emission is bad for shock revival is predicated on the assumption that there will be no re-conversion of these sterile neutrinos to active neutrino species. Indeed, our calculations suggest that such a re-conversion process could take place under some circumstances and that this re-conversion could effect an enhancement in energy and electron lepton number transport from deep in the core to the regions just below the neutrino sphere. This could {\it increase} the prospects for a viable explosion through: (1) pre-heating of the material ahead of the shock causing a reduction in the nuclear photo-disintegration burden on the shock; and (2) enhancement of the $\nu_e$ and $\bar\nu_e$ heating rate of the material under the bounce shock. 

We have found that the sterile neutrino mass and mixing parameters for which these enhancement processes can take place conform to our earlier estimates \cite{Hidaka-Fuller} of these: sterile neutrino rest mass range $1\, {\rm keV} \lesssim m_s \lesssim5\,{\rm keV}$; and  $\nu_e\rightleftharpoons\nu_s$ effective $2\times 2$ vacuum mixing angle in a range satisfying ${10}^{-10} \lesssim \sin^22\theta \lesssim {10}^{-8}$. Most significantly, we find that the neutrino mass and mixing parameter ranges which give supernova explosion enhancement include those ranges of parameters which give a possibility for viable sterile neutrino dark matter. What was missing in our earlier work \cite{Hidaka-Fuller} was an assessment of the effects of the shock itself on the neutrino forward scattering potential which governs active-sterile neutrino flavor transformation. In this paper we have done this assessment.

However, there are many uncertainties and our one-zone calculations can be regarded only as rough outlines for how active-sterile-active neutrino flavor conversion processes affect supernova core and shock physics. How can our calculations be improved on? 

First, in the context of a realistic proto-neutron star model, a self consistent hydrodynamic treatment of shock propagation coupled with active-sterile and sterile-active neutrino flavor transformation processes is in order. This could resolve tricky issues associated with the effectiveness of pre-heating in relieving the nuclear photo-dissociation burden on the shock.

Second, it would be useful to employ a detailed treatment of neutrino transport, coupled with a realistic model for the structure and equation of state of the region of the proto-neutron star near the neutrino sphere, to assess the way in which energy deposited via $\nu_e\rightarrow\nu_s\rightarrow \nu_e$ is divided up among the various active neutrino species. Also, we need to know how this deposited energy affects $Y_e$ and the emergent luminosities of the active neutrino species at and above the neutrino sphere. 

There is yet a third source of uncertainty, one which may be an issue for all core collapse supernova models. We have pointed out in this paper that the initial core bounce shock strength is an important quantity for characterizing how the shock modifies the \lq\lq fossil\rq\rq\ neutrino forward scattering potential profile which is left at the end of the core in-fall epoch. The initial shock strength depends on many factors in both the pre-collapse hydrostatic evolution epochs of the progenitor star as well as on in-fall physics issues like nuclear weak interaction rates and the sub-nuclear density equation of state. 

Ultimately, of course, the core collapse supernova problem is a grossly nonlinear one. We will have to grapple with this nonlinearity, as well as a host of fundamental nuclear physics and multi-dimensional hydrodynamic issues, if we ever hope to realize the awesome power of this \lq\lq laboratory\rq\rq\ for revealing/constraining new physics beyond the Standard Model.

\begin{acknowledgments}
This work was supported in part by NSF grant PHY-04-00359 at UCSD and
the TSI collaboration's DOE SciDAC grant at UCSD. We thank
K. Abazajian, P. Amanik, A. Kusenko, A. Mezzacappa, M. Patel, and J.~R. Wilson for valuable discussions.
\end{acknowledgments}

\bibliography{../onezone/ref_sterile_by_George}

\begin{thebibliography}{75}
\expandafter\ifx\csname natexlab\endcsname\relax\def\natexlab#1{#1}\fi
\expandafter\ifx\csname bibnamefont\endcsname\relax
  \def\bibnamefont#1{#1}\fi
\expandafter\ifx\csname bibfnamefont\endcsname\relax
  \def\bibfnamefont#1{#1}\fi
\expandafter\ifx\csname citenamefont\endcsname\relax
  \def\citenamefont#1{#1}\fi
\expandafter\ifx\csname url\endcsname\relax
  \def\url#1{\texttt{#1}}\fi
\expandafter\ifx\csname urlprefix\endcsname\relax\def\urlprefix{URL }\fi
\providecommand{\bibinfo}[2]{#2}
\providecommand{\eprint}[2][]{\url{#2}}

\bibitem[{\citenamefont{{Dodelson} and {Widrow}}(1994)}]{DM1}
\bibinfo{author}{\bibfnamefont{S.}~\bibnamefont{{Dodelson}}} \bibnamefont{and}
  \bibinfo{author}{\bibfnamefont{L.~M.} \bibnamefont{{Widrow}}},
  \bibinfo{journal}{\prl} \textbf{\bibinfo{volume}{72}}, \bibinfo{pages}{17}
  (\bibinfo{year}{1994}), \eprint{hep-ph/9303287}.

\bibitem[{\citenamefont{{Asaka} et~al.}(2006)\citenamefont{{Asaka},
  {Shaposhnikov}, and {Kusenko}}}]{DM2}
\bibinfo{author}{\bibfnamefont{T.}~\bibnamefont{{Asaka}}},
  \bibinfo{author}{\bibfnamefont{M.}~\bibnamefont{{Shaposhnikov}}},
  \bibnamefont{and}
  \bibinfo{author}{\bibfnamefont{A.}~\bibnamefont{{Kusenko}}},
  \bibinfo{journal}{Phys. Lett. B} \textbf{\bibinfo{volume}{638}},
  \bibinfo{pages}{401} (\bibinfo{year}{2006}), \eprint{hep-ph/0602150}.

\bibitem[{\citenamefont{{Shi} and {Fuller}}(1999)}]{XSF}
\bibinfo{author}{\bibfnamefont{X.}~\bibnamefont{{Shi}}} \bibnamefont{and}
  \bibinfo{author}{\bibfnamefont{G.~M.} \bibnamefont{{Fuller}}},
  \bibinfo{journal}{\prl} \textbf{\bibinfo{volume}{83}}, \bibinfo{pages}{3120}
  (\bibinfo{year}{1999}), \eprint{astro-ph/9904041}.

\bibitem[{\citenamefont{{Abazajian}
  et~al.}(2001{\natexlab{a}})\citenamefont{{Abazajian}, {Fuller}, and
  {Patel}}}]{AFP}
\bibinfo{author}{\bibfnamefont{K.}~\bibnamefont{{Abazajian}}},
  \bibinfo{author}{\bibfnamefont{G.~M.} \bibnamefont{{Fuller}}},
  \bibnamefont{and} \bibinfo{author}{\bibfnamefont{M.}~\bibnamefont{{Patel}}},
  \bibinfo{journal}{\prd} \textbf{\bibinfo{volume}{64}},
  \bibinfo{pages}{023501} (\bibinfo{year}{2001}{\natexlab{a}}),
  \eprint{astro-ph/0101524}.

\bibitem[{\citenamefont{{Dolgov} and {Hansen}}(2002)}]{DH}
\bibinfo{author}{\bibfnamefont{A.~D.} \bibnamefont{{Dolgov}}} \bibnamefont{and}
  \bibinfo{author}{\bibfnamefont{S.~H.} \bibnamefont{{Hansen}}},
  \bibinfo{journal}{Astropart. Phys.} \textbf{\bibinfo{volume}{16}},
  \bibinfo{pages}{339} (\bibinfo{year}{2002}), \eprint{hep-ph/0009083}.

\bibitem[{\citenamefont{{Abazajian} and {Fuller}}(2002)}]{AF}
\bibinfo{author}{\bibfnamefont{K.~N.} \bibnamefont{{Abazajian}}}
  \bibnamefont{and} \bibinfo{author}{\bibfnamefont{G.~M.}
  \bibnamefont{{Fuller}}}, \bibinfo{journal}{\prd}
  \textbf{\bibinfo{volume}{66}}, \bibinfo{pages}{023526}
  (\bibinfo{year}{2002}), \eprint{astro-ph/0204293}.

\bibitem[{\citenamefont{{Abazajian}}(2006{\natexlab{a}})}]{2006PhRvD..73f3506A}
\bibinfo{author}{\bibfnamefont{K.}~\bibnamefont{{Abazajian}}},
  \bibinfo{journal}{\prd} \textbf{\bibinfo{volume}{73}},
  \bibinfo{pages}{063506} (\bibinfo{year}{2006}{\natexlab{a}}),
  \eprint{astro-ph/0511630}.

\bibitem[{\citenamefont{{Biermann} and {Kusenko}}(2006)}]{BiermannKusenko}
\bibinfo{author}{\bibfnamefont{P.~L.} \bibnamefont{{Biermann}}}
  \bibnamefont{and}
  \bibinfo{author}{\bibfnamefont{A.}~\bibnamefont{{Kusenko}}},
  \bibinfo{journal}{\prl} \textbf{\bibinfo{volume}{96}},
  \bibinfo{pages}{091301} (\bibinfo{year}{2006}), \eprint{astro-ph/0601004}.

\bibitem[{\citenamefont{{Abazajian}}(2006{\natexlab{b}})}]{Kev}
\bibinfo{author}{\bibfnamefont{K.}~\bibnamefont{{Abazajian}}},
  \bibinfo{journal}{\prd} \textbf{\bibinfo{volume}{73}},
  \bibinfo{pages}{063506} (\bibinfo{year}{2006}{\natexlab{b}}),
  \eprint{astro-ph/0511630}.

\bibitem[{\citenamefont{{Abazajian} and
  {Koushiappas}}(2006)}]{AbazajianKoushiappas}
\bibinfo{author}{\bibfnamefont{K.}~\bibnamefont{{Abazajian}}} \bibnamefont{and}
  \bibinfo{author}{\bibfnamefont{S.~M.} \bibnamefont{{Koushiappas}}},
  \bibinfo{journal}{\prd} \textbf{\bibinfo{volume}{74}},
  \bibinfo{pages}{023527} (\bibinfo{year}{2006}), \eprint{astro-ph/0605271}.

\bibitem[{\citenamefont{{Shi} and {Sigl}}(1994)}]{1994PhLB..323..360S}
\bibinfo{author}{\bibfnamefont{X.}~\bibnamefont{{Shi}}} \bibnamefont{and}
  \bibinfo{author}{\bibfnamefont{G.}~\bibnamefont{{Sigl}}},
  \bibinfo{journal}{Phys. Lett. B} \textbf{\bibinfo{volume}{323}},
  \bibinfo{pages}{360} (\bibinfo{year}{1994}), \eprint{hep-ph/9312247}.

\bibitem[{\citenamefont{{Abazajian}
  et~al.}(2001{\natexlab{b}})\citenamefont{{Abazajian}, {Fuller}, and
  {Tucker}}}]{x-ray}
\bibinfo{author}{\bibfnamefont{K.}~\bibnamefont{{Abazajian}}},
  \bibinfo{author}{\bibfnamefont{G.~M.} \bibnamefont{{Fuller}}},
  \bibnamefont{and} \bibinfo{author}{\bibfnamefont{W.~H.}
  \bibnamefont{{Tucker}}}, \bibinfo{journal}{\apj}
  \textbf{\bibinfo{volume}{562}}, \bibinfo{pages}{593}
  (\bibinfo{year}{2001}{\natexlab{b}}), \eprint{astro-ph/0106002}.

\bibitem[{\citenamefont{{Boyarsky}
  et~al.}(2006{\natexlab{a}})\citenamefont{{Boyarsky}, {Neronov}, {Ruchayskiy},
  and {Shaposhnikov}}}]{x-ray_1}
\bibinfo{author}{\bibfnamefont{A.}~\bibnamefont{{Boyarsky}}},
  \bibinfo{author}{\bibfnamefont{A.}~\bibnamefont{{Neronov}}},
  \bibinfo{author}{\bibfnamefont{O.}~\bibnamefont{{Ruchayskiy}}},
  \bibnamefont{and}
  \bibinfo{author}{\bibfnamefont{M.}~\bibnamefont{{Shaposhnikov}}},
  \bibinfo{journal}{\prd} \textbf{\bibinfo{volume}{74}},
  \bibinfo{pages}{103506} (\bibinfo{year}{2006}{\natexlab{a}}),
  \eprint{arXiv:astro-ph/0603368}.

\bibitem[{\citenamefont{{Abazajian}}(2006{\natexlab{c}})}]{Kev2}
\bibinfo{author}{\bibfnamefont{K.}~\bibnamefont{{Abazajian}}},
  \bibinfo{journal}{\prd} \textbf{\bibinfo{volume}{73}},
  \bibinfo{pages}{063513} (\bibinfo{year}{2006}{\natexlab{c}}),
  \eprint{astro-ph/0512631}.

\bibitem[{\citenamefont{{Boyarsky}
  et~al.}(2006{\natexlab{b}})\citenamefont{{Boyarsky}, {Neronov}, {Ruchayskiy},
  {Shaposhnikov}, and {Tkachev}}}]{Boyarsky}
\bibinfo{author}{\bibfnamefont{A.}~\bibnamefont{{Boyarsky}}},
  \bibinfo{author}{\bibfnamefont{A.}~\bibnamefont{{Neronov}}},
  \bibinfo{author}{\bibfnamefont{O.}~\bibnamefont{{Ruchayskiy}}},
  \bibinfo{author}{\bibfnamefont{M.}~\bibnamefont{{Shaposhnikov}}},
  \bibnamefont{and}
  \bibinfo{author}{\bibfnamefont{I.}~\bibnamefont{{Tkachev}}},
  \bibinfo{journal}{Physical Review Letters} \textbf{\bibinfo{volume}{97}},
  \bibinfo{pages}{261302} (\bibinfo{year}{2006}{\natexlab{b}}),
  \eprint{arXiv:astro-ph/0603660}.

\bibitem[{\citenamefont{{Viel} et~al.}(2005)\citenamefont{{Viel},
  {Lesgourgues}, {Haehnelt}, {Matarrese}, and {Riotto}}}]{Viel}
\bibinfo{author}{\bibfnamefont{M.}~\bibnamefont{{Viel}}},
  \bibinfo{author}{\bibfnamefont{J.}~\bibnamefont{{Lesgourgues}}},
  \bibinfo{author}{\bibfnamefont{M.~G.} \bibnamefont{{Haehnelt}}},
  \bibinfo{author}{\bibfnamefont{S.}~\bibnamefont{{Matarrese}}},
  \bibnamefont{and} \bibinfo{author}{\bibfnamefont{A.}~\bibnamefont{{Riotto}}},
  \bibinfo{journal}{\prd} \textbf{\bibinfo{volume}{71}},
  \bibinfo{pages}{063534} (\bibinfo{year}{2005}), \eprint{astro-ph/0501562}.

\bibitem[{\citenamefont{{Watson} et~al.}(2006)\citenamefont{{Watson}, {Beacom},
  {Y{\"u}ksel}, and {Walker}}}]{2006PhRvD..74c3009W}
\bibinfo{author}{\bibfnamefont{C.~R.} \bibnamefont{{Watson}}},
  \bibinfo{author}{\bibfnamefont{J.~F.} \bibnamefont{{Beacom}}},
  \bibinfo{author}{\bibfnamefont{H.}~\bibnamefont{{Y{\"u}ksel}}},
  \bibnamefont{and} \bibinfo{author}{\bibfnamefont{T.~P.}
  \bibnamefont{{Walker}}}, \bibinfo{journal}{\prd}
  \textbf{\bibinfo{volume}{74}}, \bibinfo{pages}{033009}
  (\bibinfo{year}{2006}), \eprint{astro-ph/0605424}.

\bibitem[{\citenamefont{{Boyarsky}
  et~al.}(2006{\natexlab{c}})\citenamefont{{Boyarsky}, {Neronov}, {Ruchayskiy},
  and {Shaposhnikov}}}]{2006MNRAS.370..213B}
\bibinfo{author}{\bibfnamefont{A.}~\bibnamefont{{Boyarsky}}},
  \bibinfo{author}{\bibfnamefont{A.}~\bibnamefont{{Neronov}}},
  \bibinfo{author}{\bibfnamefont{O.}~\bibnamefont{{Ruchayskiy}}},
  \bibnamefont{and}
  \bibinfo{author}{\bibfnamefont{M.}~\bibnamefont{{Shaposhnikov}}},
  \bibinfo{journal}{MNRAS} \textbf{\bibinfo{volume}{370}}, \bibinfo{pages}{213}
  (\bibinfo{year}{2006}{\natexlab{c}}), \eprint{astro-ph/0512509}.

\bibitem[{\citenamefont{{Fuller} et~al.}(2003)\citenamefont{{Fuller},
  {Kusenko}, {Mocioiu}, and {Pascoli}}}]{kicks}
\bibinfo{author}{\bibfnamefont{G.~M.} \bibnamefont{{Fuller}}},
  \bibinfo{author}{\bibfnamefont{A.}~\bibnamefont{{Kusenko}}},
  \bibinfo{author}{\bibfnamefont{I.}~\bibnamefont{{Mocioiu}}},
  \bibnamefont{and}
  \bibinfo{author}{\bibfnamefont{S.}~\bibnamefont{{Pascoli}}},
  \bibinfo{journal}{\prd} \textbf{\bibinfo{volume}{68}},
  \bibinfo{pages}{103002} (\bibinfo{year}{2003}), \eprint{astro-ph/0307267}.

\bibitem[{\citenamefont{{Kusenko} and {Segre}}(1999)}]{kicks2}
\bibinfo{author}{\bibfnamefont{A.}~\bibnamefont{{Kusenko}}} \bibnamefont{and}
  \bibinfo{author}{\bibfnamefont{G.}~\bibnamefont{{Segre}}},
  \bibinfo{journal}{\prd} \textbf{\bibinfo{volume}{59}},
  \bibinfo{pages}{061302} (\bibinfo{year}{1999}), \eprint{astro-ph/9811144}.

\bibitem[{\citenamefont{{Fryer} and {Kusenko}}(2006)}]{FK}
\bibinfo{author}{\bibfnamefont{C.~L.} \bibnamefont{{Fryer}}} \bibnamefont{and}
  \bibinfo{author}{\bibfnamefont{A.}~\bibnamefont{{Kusenko}}},
  \bibinfo{journal}{Astrophys. J. Suppl.} \textbf{\bibinfo{volume}{163}},
  \bibinfo{pages}{335} (\bibinfo{year}{2006}), \eprint{astro-ph/0512033}.

\bibitem[{\citenamefont{{Athanassopoulos}
  et~al.}(1998)\citenamefont{{Athanassopoulos}, {Auerbach}, {Burman},
  {Caldwell}, {Church}, {Cohen}, {Donahue}, {Fazely}, {Federspiel}, {Garvey}
  et~al.}}]{LSND}
\bibinfo{author}{\bibfnamefont{C.}~\bibnamefont{{Athanassopoulos}}},
  \bibinfo{author}{\bibfnamefont{L.~B.} \bibnamefont{{Auerbach}}},
  \bibinfo{author}{\bibfnamefont{R.~L.} \bibnamefont{{Burman}}},
  \bibinfo{author}{\bibfnamefont{D.~O.} \bibnamefont{{Caldwell}}},
  \bibinfo{author}{\bibfnamefont{E.~D.} \bibnamefont{{Church}}},
  \bibinfo{author}{\bibfnamefont{I.}~\bibnamefont{{Cohen}}},
  \bibinfo{author}{\bibfnamefont{J.~B.} \bibnamefont{{Donahue}}},
  \bibinfo{author}{\bibfnamefont{A.}~\bibnamefont{{Fazely}}},
  \bibinfo{author}{\bibfnamefont{F.~J.} \bibnamefont{{Federspiel}}},
  \bibinfo{author}{\bibfnamefont{G.~T.} \bibnamefont{{Garvey}}},
  \bibnamefont{et~al.}, \bibinfo{journal}{\prl} \textbf{\bibinfo{volume}{81}},
  \bibinfo{pages}{1774} (\bibinfo{year}{1998}), \eprint{nucl-ex/9709006}.

\bibitem[{\citenamefont{{Sorel} et~al.}(2004)\citenamefont{{Sorel}, {Conrad},
  and {Shaevitz}}}]{LSND_1}
\bibinfo{author}{\bibfnamefont{M.}~\bibnamefont{{Sorel}}},
  \bibinfo{author}{\bibfnamefont{J.~M.} \bibnamefont{{Conrad}}},
  \bibnamefont{and} \bibinfo{author}{\bibfnamefont{M.~H.}
  \bibnamefont{{Shaevitz}}}, \bibinfo{journal}{\prd}
  \textbf{\bibinfo{volume}{70}}, \bibinfo{pages}{073004}
  (\bibinfo{year}{2004}), \eprint{hep-ph/0305255}.

\bibitem[{\citenamefont{{Aguilar} et~al.}(2001)\citenamefont{{Aguilar},
  {Auerbach}, {Burman}, {Caldwell}, {Church}, {Cochran}, {Donahue}, {Fazely},
  {Garvey}, {Gunasingha} et~al.}}]{LSND_2}
\bibinfo{author}{\bibfnamefont{A.}~\bibnamefont{{Aguilar}}},
  \bibinfo{author}{\bibfnamefont{L.~B.} \bibnamefont{{Auerbach}}},
  \bibinfo{author}{\bibfnamefont{R.~L.} \bibnamefont{{Burman}}},
  \bibinfo{author}{\bibfnamefont{D.~O.} \bibnamefont{{Caldwell}}},
  \bibinfo{author}{\bibfnamefont{E.~D.} \bibnamefont{{Church}}},
  \bibinfo{author}{\bibfnamefont{A.~K.} \bibnamefont{{Cochran}}},
  \bibinfo{author}{\bibfnamefont{J.~B.} \bibnamefont{{Donahue}}},
  \bibinfo{author}{\bibfnamefont{A.}~\bibnamefont{{Fazely}}},
  \bibinfo{author}{\bibfnamefont{G.~T.} \bibnamefont{{Garvey}}},
  \bibinfo{author}{\bibfnamefont{R.~M.} \bibnamefont{{Gunasingha}}},
  \bibnamefont{et~al.}, \bibinfo{journal}{\prd} \textbf{\bibinfo{volume}{64}},
  \bibinfo{pages}{112007} (\bibinfo{year}{2001}), \eprint{hep-ex/0104049}.

\bibitem[{\citenamefont{Aguilar-Arevalo
  et~al.}(2007)\citenamefont{Aguilar-Arevalo, Bazarko, Brice, Brown, Bugel,
  Cao, Coney, Conrad, Cox, Curioni et~al.}}]{MiniBooNE}
\bibinfo{author}{\bibfnamefont{A.~A.} \bibnamefont{Aguilar-Arevalo}},
  \bibinfo{author}{\bibfnamefont{A.~O.} \bibnamefont{Bazarko}},
  \bibinfo{author}{\bibfnamefont{S.~J.} \bibnamefont{Brice}},
  \bibinfo{author}{\bibfnamefont{B.~C.} \bibnamefont{Brown}},
  \bibinfo{author}{\bibfnamefont{L.}~\bibnamefont{Bugel}},
  \bibinfo{author}{\bibfnamefont{J.}~\bibnamefont{Cao}},
  \bibinfo{author}{\bibfnamefont{L.}~\bibnamefont{Coney}},
  \bibinfo{author}{\bibfnamefont{J.~M.} \bibnamefont{Conrad}},
  \bibinfo{author}{\bibfnamefont{D.~C.} \bibnamefont{Cox}},
  \bibinfo{author}{\bibfnamefont{A.}~\bibnamefont{Curioni}},
  \bibnamefont{et~al.} (\bibinfo{collaboration}{MiniBooNE Collaboration}),
  \bibinfo{journal}{Physical Review Letters} \textbf{\bibinfo{volume}{98}},
  \bibinfo{eid}{231801} (pages~\bibinfo{numpages}{7}) (\bibinfo{year}{2007}),
  \urlprefix\url{http://link.aps.org/abstract/PRL/v98/e231801}.

\bibitem[{\citenamefont{{Bethe} et~al.}(1979)\citenamefont{{Bethe}, {Brown},
  {Applegate}, and {Lattimer}}}]{BBAL}
\bibinfo{author}{\bibfnamefont{H.~A.} \bibnamefont{{Bethe}}},
  \bibinfo{author}{\bibfnamefont{G.~E.} \bibnamefont{{Brown}}},
  \bibinfo{author}{\bibfnamefont{J.}~\bibnamefont{{Applegate}}},
  \bibnamefont{and} \bibinfo{author}{\bibfnamefont{J.~M.}
  \bibnamefont{{Lattimer}}}, \bibinfo{journal}{Nucl. Phys. A}
  \textbf{\bibinfo{volume}{324}}, \bibinfo{pages}{487} (\bibinfo{year}{1979}).

\bibitem[{\citenamefont{{Bethe} and {Wilson}}(1985)}]{BW85}
\bibinfo{author}{\bibfnamefont{H.~A.} \bibnamefont{{Bethe}}} \bibnamefont{and}
  \bibinfo{author}{\bibfnamefont{J.~R.} \bibnamefont{{Wilson}}},
  \bibinfo{journal}{\apj} \textbf{\bibinfo{volume}{295}}, \bibinfo{pages}{14}
  (\bibinfo{year}{1985}).

\bibitem[{\citenamefont{{Burrows} et~al.}(2006)\citenamefont{{Burrows},
  {Livne}, {Dessart}, {Ott}, and {Murphy}}}]{2006ApJ...640..878B}
\bibinfo{author}{\bibfnamefont{A.}~\bibnamefont{{Burrows}}},
  \bibinfo{author}{\bibfnamefont{E.}~\bibnamefont{{Livne}}},
  \bibinfo{author}{\bibfnamefont{L.}~\bibnamefont{{Dessart}}},
  \bibinfo{author}{\bibfnamefont{C.~D.} \bibnamefont{{Ott}}}, \bibnamefont{and}
  \bibinfo{author}{\bibfnamefont{J.}~\bibnamefont{{Murphy}}},
  \bibinfo{journal}{\apj} \textbf{\bibinfo{volume}{640}}, \bibinfo{pages}{878}
  (\bibinfo{year}{2006}), \eprint{arXiv:astro-ph/0510687}.

\bibitem[{\citenamefont{{Blondin} and
  {Mezzacappa}}(2006)}]{2006ApJ...642..401B}
\bibinfo{author}{\bibfnamefont{J.~M.} \bibnamefont{{Blondin}}}
  \bibnamefont{and}
  \bibinfo{author}{\bibfnamefont{A.}~\bibnamefont{{Mezzacappa}}},
  \bibinfo{journal}{\apj} \textbf{\bibinfo{volume}{642}}, \bibinfo{pages}{401}
  (\bibinfo{year}{2006}), \eprint{astro-ph/0507181}.

\bibitem[{\citenamefont{{Fryer} and {Warren}}(2002)}]{2002ApJ...574L..65F}
\bibinfo{author}{\bibfnamefont{C.~L.} \bibnamefont{{Fryer}}} \bibnamefont{and}
  \bibinfo{author}{\bibfnamefont{M.~S.} \bibnamefont{{Warren}}},
  \bibinfo{journal}{Astrophys. J. Lett.} \textbf{\bibinfo{volume}{574}},
  \bibinfo{pages}{L65} (\bibinfo{year}{2002}), \eprint{astro-ph/0206017}.

\bibitem[{\citenamefont{{Kifonidis} et~al.}(2006)\citenamefont{{Kifonidis},
  {Plewa}, {Scheck}, {Janka}, and {M{\"u}ller}}}]{2006A&A...453..661K}
\bibinfo{author}{\bibfnamefont{K.}~\bibnamefont{{Kifonidis}}},
  \bibinfo{author}{\bibfnamefont{T.}~\bibnamefont{{Plewa}}},
  \bibinfo{author}{\bibfnamefont{L.}~\bibnamefont{{Scheck}}},
  \bibinfo{author}{\bibfnamefont{H.-T.} \bibnamefont{{Janka}}},
  \bibnamefont{and}
  \bibinfo{author}{\bibfnamefont{E.}~\bibnamefont{{M{\"u}ller}}},
  \bibinfo{journal}{Astron.Astrophys.} \textbf{\bibinfo{volume}{453}},
  \bibinfo{pages}{661} (\bibinfo{year}{2006}).

\bibitem[{\citenamefont{{Hidaka} and {Fuller}}(2006)}]{Hidaka-Fuller}
\bibinfo{author}{\bibfnamefont{J.}~\bibnamefont{{Hidaka}}} \bibnamefont{and}
  \bibinfo{author}{\bibfnamefont{G.~M.} \bibnamefont{{Fuller}}},
  \bibinfo{journal}{\prd} \textbf{\bibinfo{volume}{74}},
  \bibinfo{pages}{125015} (\bibinfo{year}{2006}),
  \eprint{arXiv:astro-ph/0609425}.

\bibitem[{\citenamefont{{Mikheyev} and {Smirnov}}(1985)}]{MSW}
\bibinfo{author}{\bibfnamefont{S.~P.} \bibnamefont{{Mikheyev}}}
  \bibnamefont{and} \bibinfo{author}{\bibfnamefont{A.~Y.}
  \bibnamefont{{Smirnov}}}, \bibinfo{journal}{Yad.~Fiz.}
  \textbf{\bibinfo{volume}{42}}, \bibinfo{pages}{1441} (\bibinfo{year}{1985}).

\bibitem[{\citenamefont{Wolfenstein}(1978)}]{MSW_1}
\bibinfo{author}{\bibfnamefont{L.}~\bibnamefont{Wolfenstein}},
  \bibinfo{journal}{Phys. Rev.} \textbf{\bibinfo{volume}{D17}},
  \bibinfo{pages}{2369} (\bibinfo{year}{1978}).

\bibitem[{\citenamefont{{Fuller} et~al.}(1987)\citenamefont{{Fuller}, {Mayle},
  {Wilson}, and {Schramm}}}]{Fuller87}
\bibinfo{author}{\bibfnamefont{G.~M.} \bibnamefont{{Fuller}}},
  \bibinfo{author}{\bibfnamefont{R.~W.} \bibnamefont{{Mayle}}},
  \bibinfo{author}{\bibfnamefont{J.~R.} \bibnamefont{{Wilson}}},
  \bibnamefont{and} \bibinfo{author}{\bibfnamefont{D.~N.}
  \bibnamefont{{Schramm}}}, \bibinfo{journal}{\apj}
  \textbf{\bibinfo{volume}{322}}, \bibinfo{pages}{795} (\bibinfo{year}{1987}).

\bibitem[{\citenamefont{{Boyanovsky} and
  {Ho}}(2007{\natexlab{a}})}]{2007PhRvD..75h5004B}
\bibinfo{author}{\bibfnamefont{D.}~\bibnamefont{{Boyanovsky}}}
  \bibnamefont{and} \bibinfo{author}{\bibfnamefont{C.~M.} \bibnamefont{{Ho}}},
  \bibinfo{journal}{\prd} \textbf{\bibinfo{volume}{75}},
  \bibinfo{pages}{085004} (\bibinfo{year}{2007}{\natexlab{a}}).

\bibitem[{\citenamefont{{Boyanovsky} and
  {Ho}}(2007{\natexlab{b}})}]{2007arXiv0705.0703B}
\bibinfo{author}{\bibfnamefont{D.}~\bibnamefont{{Boyanovsky}}}
  \bibnamefont{and} \bibinfo{author}{\bibfnamefont{C.~M.} \bibnamefont{{Ho}}},
  \bibinfo{journal}{ArXiv e-prints} \textbf{\bibinfo{volume}{705}}
  (\bibinfo{year}{2007}{\natexlab{b}}), \eprint{0705.0703}.

\bibitem[{\citenamefont{{Fuller}}(1982)}]{Fuller82}
\bibinfo{author}{\bibfnamefont{G.~M.} \bibnamefont{{Fuller}}},
  \bibinfo{journal}{\apj} \textbf{\bibinfo{volume}{252}}, \bibinfo{pages}{741}
  (\bibinfo{year}{1982}).

\bibitem[{\citenamefont{{Goldreich} and {Weber}}(1980)}]{Goldreich-Weber}
\bibinfo{author}{\bibfnamefont{P.}~\bibnamefont{{Goldreich}}} \bibnamefont{and}
  \bibinfo{author}{\bibfnamefont{S.~V.} \bibnamefont{{Weber}}},
  \bibinfo{journal}{\apj} \textbf{\bibinfo{volume}{238}}, \bibinfo{pages}{991}
  (\bibinfo{year}{1980}).

\bibitem[{\citenamefont{{Hix} et~al.}(2003{\natexlab{a}})\citenamefont{{Hix},
  {Messer}, {Mezzacappa}, {Liebend{\"o}rfer}, {Sampaio}, {Langanke}, {Dean},
  and {Mart{\'{\i}}nez-Pinedo}}}]{Hix}
\bibinfo{author}{\bibfnamefont{W.~R.} \bibnamefont{{Hix}}},
  \bibinfo{author}{\bibfnamefont{O.~E.} \bibnamefont{{Messer}}},
  \bibinfo{author}{\bibfnamefont{A.}~\bibnamefont{{Mezzacappa}}},
  \bibinfo{author}{\bibfnamefont{M.}~\bibnamefont{{Liebend{\"o}rfer}}},
  \bibinfo{author}{\bibfnamefont{J.}~\bibnamefont{{Sampaio}}},
  \bibinfo{author}{\bibfnamefont{K.}~\bibnamefont{{Langanke}}},
  \bibinfo{author}{\bibfnamefont{D.~J.} \bibnamefont{{Dean}}},
  \bibnamefont{and}
  \bibinfo{author}{\bibfnamefont{G.}~\bibnamefont{{Mart{\'{\i}}nez-Pinedo}}},
  \bibinfo{journal}{\prl} \textbf{\bibinfo{volume}{91}},
  \bibinfo{pages}{201102} (\bibinfo{year}{2003}{\natexlab{a}}),
  \eprint{astro-ph/0310883}.

\bibitem[{\citenamefont{{Hix} et~al.}(2003{\natexlab{b}})\citenamefont{{Hix},
  {Messer}, {Mezzacappa}, {Liebend{\"o}rfer}, {Sampaio}, {Langanke}, {Dean},
  and {Mart{\'{\i}}nez-Pinedo}}}]{2003PhRvL..91t1102H}
\bibinfo{author}{\bibfnamefont{W.~R.} \bibnamefont{{Hix}}},
  \bibinfo{author}{\bibfnamefont{O.~E.} \bibnamefont{{Messer}}},
  \bibinfo{author}{\bibfnamefont{A.}~\bibnamefont{{Mezzacappa}}},
  \bibinfo{author}{\bibfnamefont{M.}~\bibnamefont{{Liebend{\"o}rfer}}},
  \bibinfo{author}{\bibfnamefont{J.}~\bibnamefont{{Sampaio}}},
  \bibinfo{author}{\bibfnamefont{K.}~\bibnamefont{{Langanke}}},
  \bibinfo{author}{\bibfnamefont{D.~J.} \bibnamefont{{Dean}}},
  \bibnamefont{and}
  \bibinfo{author}{\bibfnamefont{G.}~\bibnamefont{{Mart{\'{\i}}nez-Pinedo}}},
  \bibinfo{journal}{\prl} \textbf{\bibinfo{volume}{91}},
  \bibinfo{pages}{201102} (\bibinfo{year}{2003}{\natexlab{b}}),
  \eprint{astro-ph/0310883}.

\bibitem[{\citenamefont{{Buras} et~al.}(2006)\citenamefont{{Buras}, {Rampp},
  {Janka}, and {Kifonidis}}}]{2006A&A...447.1049B}
\bibinfo{author}{\bibfnamefont{R.}~\bibnamefont{{Buras}}},
  \bibinfo{author}{\bibfnamefont{M.}~\bibnamefont{{Rampp}}},
  \bibinfo{author}{\bibfnamefont{H.-T.} \bibnamefont{{Janka}}},
  \bibnamefont{and}
  \bibinfo{author}{\bibfnamefont{K.}~\bibnamefont{{Kifonidis}}},
  \bibinfo{journal}{Astron. Astrophys.} \textbf{\bibinfo{volume}{447}},
  \bibinfo{pages}{1049} (\bibinfo{year}{2006}), \eprint{astro-ph/0507135}.

\bibitem[{\citenamefont{{Mayle} and {Wilson}}(1988)}]{1988ApJ...334..909M}
\bibinfo{author}{\bibfnamefont{R.}~\bibnamefont{{Mayle}}} \bibnamefont{and}
  \bibinfo{author}{\bibfnamefont{J.~R.} \bibnamefont{{Wilson}}},
  \bibinfo{journal}{\apj} \textbf{\bibinfo{volume}{334}}, \bibinfo{pages}{909}
  (\bibinfo{year}{1988}).

\bibitem[{\citenamefont{{Wilson} and {Mayle}}(1993)}]{1993PhR...227...97W}
\bibinfo{author}{\bibfnamefont{J.~R.} \bibnamefont{{Wilson}}} \bibnamefont{and}
  \bibinfo{author}{\bibfnamefont{R.~W.} \bibnamefont{{Mayle}}},
  \bibinfo{journal}{Phys. Rep.} \textbf{\bibinfo{volume}{227}},
  \bibinfo{pages}{97} (\bibinfo{year}{1993}).

\bibitem[{\citenamefont{{Liebend{\"o}rfer}
  et~al.}(2004)\citenamefont{{Liebend{\"o}rfer}, {Messer}, {Mezzacappa},
  {Bruenn}, {Cardall}, and {Thielemann}}}]{2004ApJS..150..263L}
\bibinfo{author}{\bibfnamefont{M.}~\bibnamefont{{Liebend{\"o}rfer}}},
  \bibinfo{author}{\bibfnamefont{O.~E.~B.} \bibnamefont{{Messer}}},
  \bibinfo{author}{\bibfnamefont{A.}~\bibnamefont{{Mezzacappa}}},
  \bibinfo{author}{\bibfnamefont{S.~W.} \bibnamefont{{Bruenn}}},
  \bibinfo{author}{\bibfnamefont{C.~Y.} \bibnamefont{{Cardall}}},
  \bibnamefont{and} \bibinfo{author}{\bibfnamefont{F.-K.}
  \bibnamefont{{Thielemann}}}, \bibinfo{journal}{Astrophys. J. Suppl.}
  \textbf{\bibinfo{volume}{150}}, \bibinfo{pages}{263} (\bibinfo{year}{2004}),
  \eprint{astro-ph/0207036}.

\bibitem[{\citenamefont{{Thompson} et~al.}(2003)\citenamefont{{Thompson},
  {Burrows}, and {Pinto}}}]{2003ApJ...592..434T}
\bibinfo{author}{\bibfnamefont{T.~A.} \bibnamefont{{Thompson}}},
  \bibinfo{author}{\bibfnamefont{A.}~\bibnamefont{{Burrows}}},
  \bibnamefont{and} \bibinfo{author}{\bibfnamefont{P.~A.}
  \bibnamefont{{Pinto}}}, \bibinfo{journal}{\apj}
  \textbf{\bibinfo{volume}{592}}, \bibinfo{pages}{434} (\bibinfo{year}{2003}),
  \eprint{astro-ph/0211194}.

\bibitem[{\citenamefont{{Liebend{\"o}rfer}
  et~al.}(2005)\citenamefont{{Liebend{\"o}rfer}, {Rampp}, {Janka}, and
  {Mezzacappa}}}]{2005ApJ...620..840L}
\bibinfo{author}{\bibfnamefont{M.}~\bibnamefont{{Liebend{\"o}rfer}}},
  \bibinfo{author}{\bibfnamefont{M.}~\bibnamefont{{Rampp}}},
  \bibinfo{author}{\bibfnamefont{H.-T.} \bibnamefont{{Janka}}},
  \bibnamefont{and}
  \bibinfo{author}{\bibfnamefont{A.}~\bibnamefont{{Mezzacappa}}},
  \bibinfo{journal}{\apj} \textbf{\bibinfo{volume}{620}}, \bibinfo{pages}{840}
  (\bibinfo{year}{2005}), \eprint{astro-ph/0310662}.

\bibitem[{\citenamefont{{Mezzacappa} and
  {Blondin}}(2005)}]{2005AAS...207.1701M}
\bibinfo{author}{\bibfnamefont{A.}~\bibnamefont{{Mezzacappa}}}
  \bibnamefont{and} \bibinfo{author}{\bibfnamefont{J.~M.}
  \bibnamefont{{Blondin}}}, \bibinfo{journal}{Bulletin of the American
  Astronomical Society} \textbf{\bibinfo{volume}{37}}, \bibinfo{pages}{1181}
  (\bibinfo{year}{2005}).

\bibitem[{\citenamefont{{Mezzacappa} et~al.}(2004)\citenamefont{{Mezzacappa},
  {Liebend{\"o}rfer}, {Cardall}, {Messer}, and {Bruenn}}}]{2004rpao.conf..224L}
\bibinfo{author}{\bibfnamefont{A.}~\bibnamefont{{Mezzacappa}}},
  \bibinfo{author}{\bibfnamefont{M.}~\bibnamefont{{Liebend{\"o}rfer}}},
  \bibinfo{author}{\bibfnamefont{C.~Y.} \bibnamefont{{Cardall}}},
  \bibinfo{author}{\bibfnamefont{O.~E.~B.} \bibnamefont{{Messer}}},
  \bibnamefont{and} \bibinfo{author}{\bibfnamefont{S.~W.}
  \bibnamefont{{Bruenn}}}, in \emph{\bibinfo{booktitle}{Stellar Collapse}},
  edited by \bibinfo{editor}{\bibfnamefont{C.~L.} \bibnamefont{{Fryer}}}
  (\bibinfo{publisher}{Kluwer}, \bibinfo{year}{2004}), p.~\bibinfo{pages}{99}.

\bibitem[{\citenamefont{{Walder} et~al.}(2005)\citenamefont{{Walder},
  {Burrows}, {Ott}, {Livne}, {Lichtenstadt}, and
  {Jarrah}}}]{2005ApJ...626..317W}
\bibinfo{author}{\bibfnamefont{R.}~\bibnamefont{{Walder}}},
  \bibinfo{author}{\bibfnamefont{A.}~\bibnamefont{{Burrows}}},
  \bibinfo{author}{\bibfnamefont{C.~D.} \bibnamefont{{Ott}}},
  \bibinfo{author}{\bibfnamefont{E.}~\bibnamefont{{Livne}}},
  \bibinfo{author}{\bibfnamefont{I.}~\bibnamefont{{Lichtenstadt}}},
  \bibnamefont{and} \bibinfo{author}{\bibfnamefont{M.}~\bibnamefont{{Jarrah}}},
  \bibinfo{journal}{\apj} \textbf{\bibinfo{volume}{626}}, \bibinfo{pages}{317}
  (\bibinfo{year}{2005}), \eprint{astro-ph/0412187}.

\bibitem[{\citenamefont{{Cardall} et~al.}(2005)\citenamefont{{Cardall},
  {Lentz}, and {Mezzacappa}}}]{2005PhRvD..72d3007C}
\bibinfo{author}{\bibfnamefont{C.~Y.} \bibnamefont{{Cardall}}},
  \bibinfo{author}{\bibfnamefont{E.~J.} \bibnamefont{{Lentz}}},
  \bibnamefont{and}
  \bibinfo{author}{\bibfnamefont{A.}~\bibnamefont{{Mezzacappa}}},
  \bibinfo{journal}{\prd} \textbf{\bibinfo{volume}{72}},
  \bibinfo{pages}{043007} (\bibinfo{year}{2005}), \eprint{astro-ph/0510702}.

\bibitem[{\citenamefont{{Bruenn} et~al.}(2001)\citenamefont{{Bruenn}, {De
  Nisco}, and {Mezzacappa}}}]{2001ApJ...560..326B}
\bibinfo{author}{\bibfnamefont{S.~W.} \bibnamefont{{Bruenn}}},
  \bibinfo{author}{\bibfnamefont{K.~R.} \bibnamefont{{De Nisco}}},
  \bibnamefont{and}
  \bibinfo{author}{\bibfnamefont{A.}~\bibnamefont{{Mezzacappa}}},
  \bibinfo{journal}{\apj} \textbf{\bibinfo{volume}{560}}, \bibinfo{pages}{326}
  (\bibinfo{year}{2001}), \eprint{astro-ph/0101400}.

\bibitem[{\citenamefont{{Swesty} and {Myra}}(2006)}]{2006astro.ph..7281S}
\bibinfo{author}{\bibfnamefont{F.~D.} \bibnamefont{{Swesty}}} \bibnamefont{and}
  \bibinfo{author}{\bibfnamefont{E.~S.} \bibnamefont{{Myra}}},
  \bibinfo{journal}{ArXiv Astrophysics e-prints}  (\bibinfo{year}{2006}),
  \eprint{astro-ph/0607281}.

\bibitem[{\citenamefont{{Abazajian}
  et~al.}(2001{\natexlab{c}})\citenamefont{{Abazajian}, {Fuller}, and
  {Patel}}}]{2001PhRvD..64b3501A}
\bibinfo{author}{\bibfnamefont{K.}~\bibnamefont{{Abazajian}}},
  \bibinfo{author}{\bibfnamefont{G.~M.} \bibnamefont{{Fuller}}},
  \bibnamefont{and} \bibinfo{author}{\bibfnamefont{M.}~\bibnamefont{{Patel}}},
  \bibinfo{journal}{\prd} \textbf{\bibinfo{volume}{64}},
  \bibinfo{pages}{023501} (\bibinfo{year}{2001}{\natexlab{c}}),
  \eprint{astro-ph/0101524}.

\bibitem[{\citenamefont{{Kainulainen} et~al.}(1991)\citenamefont{{Kainulainen},
  {Maalampi}, and {Peltoniemi}}}]{1991NuPhB.358..435K}
\bibinfo{author}{\bibfnamefont{K.}~\bibnamefont{{Kainulainen}}},
  \bibinfo{author}{\bibfnamefont{J.}~\bibnamefont{{Maalampi}}},
  \bibnamefont{and} \bibinfo{author}{\bibfnamefont{J.~T.}
  \bibnamefont{{Peltoniemi}}}, \bibinfo{journal}{Nucl. Phys. B}
  \textbf{\bibinfo{volume}{358}}, \bibinfo{pages}{435} (\bibinfo{year}{1991}).

\bibitem[{\citenamefont{{Raffelt} and {Sigl}}(1993)}]{1993APh.....1..165R}
\bibinfo{author}{\bibfnamefont{G.}~\bibnamefont{{Raffelt}}} \bibnamefont{and}
  \bibinfo{author}{\bibfnamefont{G.}~\bibnamefont{{Sigl}}},
  \bibinfo{journal}{Astropart. Phys.} \textbf{\bibinfo{volume}{1}},
  \bibinfo{pages}{165} (\bibinfo{year}{1993}), \eprint{astro-ph/9209005}.

\bibitem[{\citenamefont{{Raffelt}}(1996)}]{1996slfp.book.....R}
\bibinfo{author}{\bibfnamefont{G.}~\bibnamefont{{Raffelt}}},
  \emph{\bibinfo{title}{{Stars as laboratories for fundamental physics : the
  astrophysics of neutrinos, axions, and other weakly interacting particles}}}
  (\bibinfo{publisher}{University of Chicago Press}, \bibinfo{year}{1996}).

\bibitem[{\citenamefont{{Pons} et~al.}(1999)\citenamefont{{Pons}, {Reddy},
  {Prakash}, {Lattimer}, and {Miralles}}}]{1999ApJ...513..780P}
\bibinfo{author}{\bibfnamefont{J.~A.} \bibnamefont{{Pons}}},
  \bibinfo{author}{\bibfnamefont{S.}~\bibnamefont{{Reddy}}},
  \bibinfo{author}{\bibfnamefont{M.}~\bibnamefont{{Prakash}}},
  \bibinfo{author}{\bibfnamefont{J.~M.} \bibnamefont{{Lattimer}}},
  \bibnamefont{and} \bibinfo{author}{\bibfnamefont{J.~A.}
  \bibnamefont{{Miralles}}}, \bibinfo{journal}{\apj}
  \textbf{\bibinfo{volume}{513}}, \bibinfo{pages}{780} (\bibinfo{year}{1999}),
  \eprint{astro-ph/9807040}.

\bibitem[{\citenamefont{{Qian} et~al.}(1993)\citenamefont{{Qian}, {Fuller},
  {Mathews}, {Mayle}, {Wilson}, and {Woosley}}}]{Qian}
\bibinfo{author}{\bibfnamefont{Y.-Z.} \bibnamefont{{Qian}}},
  \bibinfo{author}{\bibfnamefont{G.~M.} \bibnamefont{{Fuller}}},
  \bibinfo{author}{\bibfnamefont{G.~J.} \bibnamefont{{Mathews}}},
  \bibinfo{author}{\bibfnamefont{R.~W.} \bibnamefont{{Mayle}}},
  \bibinfo{author}{\bibfnamefont{J.~R.} \bibnamefont{{Wilson}}},
  \bibnamefont{and} \bibinfo{author}{\bibfnamefont{S.~E.}
  \bibnamefont{{Woosley}}}, \bibinfo{journal}{\prl}
  \textbf{\bibinfo{volume}{71}}, \bibinfo{pages}{1965} (\bibinfo{year}{1993}).

\bibitem[{\citenamefont{{Woosley} et~al.}(1994)\citenamefont{{Woosley},
  {Wilson}, {Mathews}, {Hoffman}, and {Meyer}}}]{1994ApJ...433..229W}
\bibinfo{author}{\bibfnamefont{S.~E.} \bibnamefont{{Woosley}}},
  \bibinfo{author}{\bibfnamefont{J.~R.} \bibnamefont{{Wilson}}},
  \bibinfo{author}{\bibfnamefont{G.~J.} \bibnamefont{{Mathews}}},
  \bibinfo{author}{\bibfnamefont{R.~D.} \bibnamefont{{Hoffman}}},
  \bibnamefont{and} \bibinfo{author}{\bibfnamefont{B.~S.}
  \bibnamefont{{Meyer}}}, \bibinfo{journal}{\apj}
  \textbf{\bibinfo{volume}{433}}, \bibinfo{pages}{229} (\bibinfo{year}{1994}).

\bibitem[{\citenamefont{{Hoffman} et~al.}(1996)\citenamefont{{Hoffman},
  {Woosley}, {Fuller}, and {Meyer}}}]{1996ApJ...460..478H}
\bibinfo{author}{\bibfnamefont{R.~D.} \bibnamefont{{Hoffman}}},
  \bibinfo{author}{\bibfnamefont{S.~E.} \bibnamefont{{Woosley}}},
  \bibinfo{author}{\bibfnamefont{G.~M.} \bibnamefont{{Fuller}}},
  \bibnamefont{and} \bibinfo{author}{\bibfnamefont{B.~S.}
  \bibnamefont{{Meyer}}}, \bibinfo{journal}{\apj}
  \textbf{\bibinfo{volume}{460}}, \bibinfo{pages}{478} (\bibinfo{year}{1996}).

\bibitem[{\citenamefont{{McLaughlin} et~al.}(1999)\citenamefont{{McLaughlin},
  {Fetter}, {Balantekin}, and {Fuller}}}]{1999PhRvC..59.2873M}
\bibinfo{author}{\bibfnamefont{G.~C.} \bibnamefont{{McLaughlin}}},
  \bibinfo{author}{\bibfnamefont{J.~M.} \bibnamefont{{Fetter}}},
  \bibinfo{author}{\bibfnamefont{A.~B.} \bibnamefont{{Balantekin}}},
  \bibnamefont{and} \bibinfo{author}{\bibfnamefont{G.~M.}
  \bibnamefont{{Fuller}}}, \bibinfo{journal}{\prc}
  \textbf{\bibinfo{volume}{59}}, \bibinfo{pages}{2873} (\bibinfo{year}{1999}),
  \eprint{astro-ph/9902106}.

\bibitem[{\citenamefont{Fetter et~al.}(2003)\citenamefont{Fetter, McLaughlin,
  Balantekin, and Fuller}}]{Fetter:2002xx}
\bibinfo{author}{\bibfnamefont{J.}~\bibnamefont{Fetter}},
  \bibinfo{author}{\bibfnamefont{G.~C.} \bibnamefont{McLaughlin}},
  \bibinfo{author}{\bibfnamefont{A.~B.} \bibnamefont{Balantekin}},
  \bibnamefont{and} \bibinfo{author}{\bibfnamefont{G.~M.}
  \bibnamefont{Fuller}}, \bibinfo{journal}{Astropart. Phys.}
  \textbf{\bibinfo{volume}{18}}, \bibinfo{pages}{433} (\bibinfo{year}{2003}),
  \eprint{hep-ph/0205029}.

\bibitem[{\citenamefont{{N{\"o}tzold} and
  {Raffelt}}(1988)}]{1988NuPhB.307..924N}
\bibinfo{author}{\bibfnamefont{D.}~\bibnamefont{{N{\"o}tzold}}}
  \bibnamefont{and}
  \bibinfo{author}{\bibfnamefont{G.}~\bibnamefont{{Raffelt}}},
  \bibinfo{journal}{Nuclear Physics B} \textbf{\bibinfo{volume}{307}},
  \bibinfo{pages}{924} (\bibinfo{year}{1988}).

\bibitem[{\citenamefont{{Pantaleone}}(1992)}]{1992PhRvD..46..510P}
\bibinfo{author}{\bibfnamefont{J.}~\bibnamefont{{Pantaleone}}},
  \bibinfo{journal}{\prd} \textbf{\bibinfo{volume}{46}}, \bibinfo{pages}{510}
  (\bibinfo{year}{1992}).

\bibitem[{\citenamefont{{Sigl} and {Raffelt}}(1993)}]{1993NuPhB.406..423S}
\bibinfo{author}{\bibfnamefont{G.}~\bibnamefont{{Sigl}}} \bibnamefont{and}
  \bibinfo{author}{\bibfnamefont{G.}~\bibnamefont{{Raffelt}}},
  \bibinfo{journal}{Nuclear Physics B} \textbf{\bibinfo{volume}{406}},
  \bibinfo{pages}{423} (\bibinfo{year}{1993}).

\bibitem[{\citenamefont{{Fuller} et~al.}(1992)\citenamefont{{Fuller}, {Mayle},
  {Meyer}, and {Wilson}}}]{FM}
\bibinfo{author}{\bibfnamefont{G.~M.} \bibnamefont{{Fuller}}},
  \bibinfo{author}{\bibfnamefont{R.}~\bibnamefont{{Mayle}}},
  \bibinfo{author}{\bibfnamefont{B.~S.} \bibnamefont{{Meyer}}},
  \bibnamefont{and} \bibinfo{author}{\bibfnamefont{J.~R.}
  \bibnamefont{{Wilson}}}, \bibinfo{journal}{\apj}
  \textbf{\bibinfo{volume}{389}}, \bibinfo{pages}{517} (\bibinfo{year}{1992}).

\bibitem[{\citenamefont{{Pastor} and {Raffelt}}(2002)}]{PastorRaffelt}
\bibinfo{author}{\bibfnamefont{S.}~\bibnamefont{{Pastor}}} \bibnamefont{and}
  \bibinfo{author}{\bibfnamefont{G.}~\bibnamefont{{Raffelt}}},
  \bibinfo{journal}{\prl} \textbf{\bibinfo{volume}{89}},
  \bibinfo{pages}{191101} (\bibinfo{year}{2002}), \eprint{astro-ph/0207281}.

\bibitem[{\citenamefont{{Balantekin} and
  {Y{\"u}ksel}}(2005)}]{2005NJPh....7...51B}
\bibinfo{author}{\bibfnamefont{A.~B.} \bibnamefont{{Balantekin}}}
  \bibnamefont{and}
  \bibinfo{author}{\bibfnamefont{H.}~\bibnamefont{{Y{\"u}ksel}}},
  \bibinfo{journal}{New Journal of Physics} \textbf{\bibinfo{volume}{7}},
  \bibinfo{pages}{51} (\bibinfo{year}{2005}), \eprint{arXiv:astro-ph/0411159}.

\bibitem[{\citenamefont{{Fuller} and {Qian}}(2006)}]{2006PhRvD..73b3004F}
\bibinfo{author}{\bibfnamefont{G.~M.} \bibnamefont{{Fuller}}} \bibnamefont{and}
  \bibinfo{author}{\bibfnamefont{Y.-Z.} \bibnamefont{{Qian}}},
  \bibinfo{journal}{\prd} \textbf{\bibinfo{volume}{73}},
  \bibinfo{pages}{023004} (\bibinfo{year}{2006}),
  \eprint{arXiv:astro-ph/0505240}.

\bibitem[{\citenamefont{{Hannestad} et~al.}(2006)\citenamefont{{Hannestad},
  {Raffelt}, {Sigl}, and {Wong}}}]{2006PhRvD..74j5010H}
\bibinfo{author}{\bibfnamefont{S.}~\bibnamefont{{Hannestad}}},
  \bibinfo{author}{\bibfnamefont{G.~G.} \bibnamefont{{Raffelt}}},
  \bibinfo{author}{\bibfnamefont{G.}~\bibnamefont{{Sigl}}}, \bibnamefont{and}
  \bibinfo{author}{\bibfnamefont{Y.~Y.~Y.} \bibnamefont{{Wong}}},
  \bibinfo{journal}{\prd} \textbf{\bibinfo{volume}{74}},
  \bibinfo{pages}{105010} (\bibinfo{year}{2006}),
  \eprint{arXiv:astro-ph/0608695}.

\bibitem[{\citenamefont{{Duan} et~al.}(2006{\natexlab{a}})\citenamefont{{Duan},
  {Fuller}, and {Qian}}}]{2006PhRvD..74l3004D}
\bibinfo{author}{\bibfnamefont{H.}~\bibnamefont{{Duan}}},
  \bibinfo{author}{\bibfnamefont{G.~M.} \bibnamefont{{Fuller}}},
  \bibnamefont{and} \bibinfo{author}{\bibfnamefont{Y.-Z.}
  \bibnamefont{{Qian}}}, \bibinfo{journal}{\prd} \textbf{\bibinfo{volume}{74}},
  \bibinfo{pages}{123004} (\bibinfo{year}{2006}{\natexlab{a}}),
  \eprint{arXiv:astro-ph/0511275}.

\bibitem[{\citenamefont{{Duan} et~al.}(2006{\natexlab{b}})\citenamefont{{Duan},
  {Fuller}, {Carlson}, and {Qian}}}]{2006PhRvD..74j5014D}
\bibinfo{author}{\bibfnamefont{H.}~\bibnamefont{{Duan}}},
  \bibinfo{author}{\bibfnamefont{G.~M.} \bibnamefont{{Fuller}}},
  \bibinfo{author}{\bibfnamefont{J.}~\bibnamefont{{Carlson}}},
  \bibnamefont{and} \bibinfo{author}{\bibfnamefont{Y.-Z.}
  \bibnamefont{{Qian}}}, \bibinfo{journal}{\prd} \textbf{\bibinfo{volume}{74}},
  \bibinfo{pages}{105014} (\bibinfo{year}{2006}{\natexlab{b}}),
  \eprint{arXiv:astro-ph/0606616}.

\bibitem[{\citenamefont{{Duan} et~al.}(2006{\natexlab{c}})\citenamefont{{Duan},
  {Fuller}, {Carlson}, and {Qian}}}]{2006PhRvL..97x1101D}
\bibinfo{author}{\bibfnamefont{H.}~\bibnamefont{{Duan}}},
  \bibinfo{author}{\bibfnamefont{G.~M.} \bibnamefont{{Fuller}}},
  \bibinfo{author}{\bibfnamefont{J.}~\bibnamefont{{Carlson}}},
  \bibnamefont{and} \bibinfo{author}{\bibfnamefont{Y.-Z.}
  \bibnamefont{{Qian}}}, \bibinfo{journal}{Physical Review Letters}
  \textbf{\bibinfo{volume}{97}}, \bibinfo{pages}{241101}
  (\bibinfo{year}{2006}{\natexlab{c}}), \eprint{arXiv:astro-ph/0608050}.

\bibitem[{\citenamefont{Duan et~al.}(2007)\citenamefont{Duan, Fuller, Carlson,
  and Qian}}]{duan:125005}
\bibinfo{author}{\bibfnamefont{H.}~\bibnamefont{Duan}},
  \bibinfo{author}{\bibfnamefont{G.~M.} \bibnamefont{Fuller}},
  \bibinfo{author}{\bibfnamefont{J.}~\bibnamefont{Carlson}}, \bibnamefont{and}
  \bibinfo{author}{\bibfnamefont{Y.-Z.} \bibnamefont{Qian}},
  \bibinfo{journal}{Physical Review D (Particles, Fields, Gravitation, and
  Cosmology)} \textbf{\bibinfo{volume}{75}}, \bibinfo{eid}{125005}
  (pages~\bibinfo{numpages}{17}) (\bibinfo{year}{2007}),
  \urlprefix\url{http://link.aps.org/abstract/PRD/v75/e125005}.

\end{thebibliography}

\end{document}